\documentclass[12pt]{iopart}
\usepackage[pdftex]{graphicx}
\usepackage{amssymb}
\begin{document}

\title{Dynamics of the Aharonov-Bohm effect}

\author{Neven Simicevic \footnote[3]{Correspondence should be addressed to Louisiana Tech University, 
PO Box 10348, Ruston, LA 71272, Tel: +1.318.257.3591, Fax: +1.318.257.4228, 
E-mail: neven@phys.latech.edu}}

\address{\ Center for Applied Physics Studies, Louisiana Tech University,
 Ruston, LA 71272, USA}

\begin{abstract}

The time-dependent Dirac equation is solved using the three-dimensional
Finite Difference-Time Domain (FDTD) method. The dynamics of the electron wave packet 
in a vector potential is studied in the arrangements associated with the Aharonov-Bohm effect.
The solution of the Dirac equation showed a change in the velocity of the electron wave packet 
even in a region where no fields of the unperturbed solenoid acted on the electron. The solution 
of the Dirac equation qualitatively agreed with the prediction of classical dynamics under the assumption 
that the dynamics was defined by the conservation of generalized or canonical momentum of the electron.

\end{abstract}


\maketitle

\section{Introduction}

The Finite Difference-Time Domain (FDTD) method, 
originally introduced by Kane Yee \cite{Yee66} to solve Maxwell's equations, is for the first time 
applied to solve the time-dependent three-dimensional Dirac equation. The $Zitterbewegung$ and 
the dynamics of well-localized electron were used as examples of FDTD 
applied to the case of free electrons \cite{Sim08}. The motion of electron wave packets 
inside and scattering from the potential step barrier or linearly dependent potential,
arrangements associated with the Klein paradox \cite{Klein29},  were
used as examples of electron interaction with an electromagnetic scalar potential \cite{Sim09}. 
In this paper, a FDTD study of  the dynamics 
of an electron wave packet under the influence of a vector potential is presented. Such a dynamic behavior 
is most often associated with the Aharonov-Bohm effect \cite{AB59}.

In the Aharonov-Bohm effect, as a manifestation of quantum mechanics, charged particles passing around a long solenoid 
can feel a magnetic flux even when all the fields of the unperturbed solenoid are zero in the 
region through which the particles travel. The shifts in the phase of the wave functions 
describing the particles have been experimentally verified by its effect on the interference 
fringes \cite {Cham60,Tono86,Osak86,Pesh89}. Since for the unperturbed solenoid, there are no
classical forces acting on the charged particles in the zero field region, the theoretical 
description of the Aharonov-Bohm effect contains a number of assumptions. They include
assumptions on nonlocal features of quantum mechanics, the physical meaning of the vector potential,
topological effects, etc., but generally accepted physical understanding is still lacking \cite{Hege08}.
While there is still an open question on the presence of classical forces responsible for 
the Aharonov-Bohm effect \cite{Hege08,Boy06,Boy08}, on a macroscopic level they have 
not been observed \cite{Cap07}. 

Proper quantum-mechanical description of the dynamics of a relativistic charged particle involves 
the solution of the Dirac equation in the time domain. In addition to initial conditions, such a dynamics 
is defined only by the configuration of the electromagnetic scalar and vector potentials, and does not involve knowledge 
or any assumption on ``classical forces". The solutions of the time-dependent Dirac equation,
some of which are presented in this paper, can shed light and fill critical knowledge gaps on the theoretical
and experimental interpretations of the mechanism of the Aharonov-Bohm effect, including the existence
or non-existence of classical forces.

\section{Time-dependent solution of the Dirac equation}

The FDTD solutions of the time-dependent Dirac equation were obtained 
for the case when the
electromagnetic field described by the four-potential ${A^{\mu}=\{A_{0}(x),\vec A(x)\}}$
was minimally coupled to the particle \cite{Grein85,Sak87}
\begin{equation}
{\imath \hbar {\frac{\partial \Psi}{\partial t}}= ({H}_{free}+{H}_{int}) \Psi},
\label{Dirac_eq}
\end{equation}
where
\begin{equation}
{{H}_{free} = -\imath c\hbar {{\bf\alpha} \cdot \nabla} + \beta m c^{2}},
\end{equation}
\begin{equation}
{{H}_{int} = - e {{\bf\alpha} \cdot {\vec A} } + e A_{0}},
\end{equation}
and
\begin{equation}
{\Psi (x) =\left( \begin{array} {c} \Psi_{1} (x) \\ \Psi_{2} (x)
\\ \Psi_{3} (x)\\ \Psi_{4} (x) \end{array} \right)}.
\end{equation}
The matrices ${\bf\alpha}$ and $\beta$ were expressed using $2 \times 2$ Pauli
matrices $\bf\sigma^{'}s$ and the $2 \times 2$ unit matrix $I$.

In the FDTD method, the time dependent solution of the Dirac equation was obtained 
using updating difference equations \cite{Sim08}. As an example, the values of $\Psi_{1}$ at 
the position $(i\Delta x,j\Delta y,k\Delta z)$ and at the time step $(n+1/2)\Delta t$ were obtained 
using the equation

\begin{eqnarray}
 \Psi_{1}^{n+1/2}(I,J,K)&=&\frac{2-C^{n}(I,J,K)}{C^{n}(I,J,K)}\Psi_{1}^{n-1/2}(I,J,K) \nonumber \\
&-&{\frac{c\Delta t}{2\Delta x C^{n}(I,J,K)}}
[ \Psi_{3}^{n}(I,J,K+1)-\Psi_{3}^{n}(I,J,K-1) \nonumber \\
&+&\Psi_{4}^{n}(I+1,J,K)-\Psi_{4}^{n}(I-1,J,K)-i(\Psi_{4}^{n}(I,J+1,K) \nonumber \\
&-&\Psi_{4}^{n}(I,J-1,K))]
+i{\frac{e\Delta t}{\hbar C^{n}(I,J,K)}}[ A^{n}_{1}(I,J,K)\Psi_{4}^{n}(I,J,K) \nonumber \\
&-&iA^{n}_{2}(I,J,K)\Psi_{4}^{n}(I,J,K)
+A^{n}_{3}(I,J,K)\Psi_{3}^{n}(I,J,K)],
\label{Psi_1}
\end{eqnarray}
where $C^{n}(I,J,K)=1+i\frac{\Delta t}{2\hbar}[mc^{2}+eA^{n}_{0}(I,J,K)]$.
The space and time were discretized using uniform rectangular lattices of size
$\Delta x$, $\Delta y$ and $\Delta z$, and uniform time increment $\Delta t$.
While it is not generally required, in the Eq. (\ref{Psi_1}) $\Delta x = \Delta y = \Delta z$. 
Updating equations for $\Psi_{2}$, $\Psi_{3}$, and $\Psi_{4}$ were
constructed in a similar way. as a result, the dynamics of a Dirac electron can be studied
in any environment described by a four-potential $A^{\mu}$ regardless
of its complexity and time dependency. In this paper we studied the dynamics defined
by the vector potential $\vec A(x) \neq 0$.

The Dirac equation is a differential equation of the first order and linear
in $\partial / \partial t$. As in the case of Maxwell's equations, the entire 
dynamics of the electron is defined, only by its initial wave function. In this paper, the dynamics
of a wave packet was defined by the initial wave function of the form
\begin{equation}
{\Psi (\vec x,0) =N \sqrt{\frac{E+mc^{2}}{2E}}\left( \begin{array} {c} 1 \\ 0
\\ \frac{p_{3}c}{E+mc^{2}}\\ \frac{(p_{1}+ip_{2})c}{E+mc^{2}}\end{array} \right)}
e^{-\frac{\vec x \cdot \vec x }{4x_{0}^{2}}+\frac{i\vec p \cdot \vec x}{\hbar}},
\label{Wave_packet}
\end{equation}
where $N=[(2\pi)^{3/2}x_{0}^{3}]^{-1/2}$ is a normalizing constant. Eq. (\ref{Wave_packet})
represents a wave packet whose initial probability distribution is of a normalized Gaussian
shape. Its size is defined by the constant $x_{0}$, its spin is pointed along the z-axis, and its
motion is defined by the values of the momenta $p_{1},p_{2}$, and $p_{3}$. Some consequences of the 
initial localization of the wave packet on the overall dynamics of the electron were studied 
in Ref. \cite{Sim08,Huang52}.

\section{Validation of the computation}
\subsection{Dynamics of a wave packet in a strong uniform magnetic field}

The dynamics of a wave packet is very complex. The dynamics of a particle described 
by the wave packet in Eq. (\ref{Wave_packet}) depends on its localization, defined by 
the Gaussian component of the wave function, and its initial momentum, 
part of the wave function's phase. While an extensive study was done on the dynamics of
the wave packet related to the scalar component $A_{0}(x)$ of a four-potential \cite{Sim09}, 
such a study does not validate the dynamics of the wave packet related to the 
vector component of a four-potential, $\vec A(x)$.

Applying the FDTD method to study the dynamics of a wave packet in a vector potential
associated with a strong uniform magnetic field is not difficult. Classically, the dynamics consist of
uniform rotational motion. 
In the relativistic quantum-mechanical description, however, even such a simple dynamics 
can validate the FDTD computation only up to a certain level. 
As pointed out in Ref. \cite{Schl08} "the dynamics is particularly
rich and not adequately described by semiclassical approximations". In Ref. \cite{Demi08} 
it was demonstrated that in the presence of an external magnetic field the wave packet splits into two parts 
which rotate with different cyclotron frequencies, and after a few periods, the motion acquires irregular
character.  As a result of such a complex dynamics, when comparing the results obtained by
the FDTD method and the computation described in Ref. \cite{Schl08} and \cite{Demi08},
we could not expect more than a qualitative agreement. In addition to this qualitative agreement,
of equal importance to the validation of the computation is the consistency of the results obtained 
for different vector potential gauges.

In this paper, the motion of the wave packet described by Eq. (\ref{Wave_packet}) 
was studied in a uniform magnetic field oriented along the y-axis

\begin{equation}
\vec B=(0,B_{0},0).
\label{Uni_mag_field}
\end{equation}
For this field the corresponding vector potential in a rotationally invariant gauge is
\begin{equation}
\vec A={B_{0} \over 2} (z,0,-x).
\label{Uni_mag_potent}
\end{equation}
The dynamics of the wave packet in a uniform magnetic field was first 
obtained by solving the Dirac equation for this vector potential.
 
\begin{figure}
\centering
{\scalebox{.7}{\includegraphics{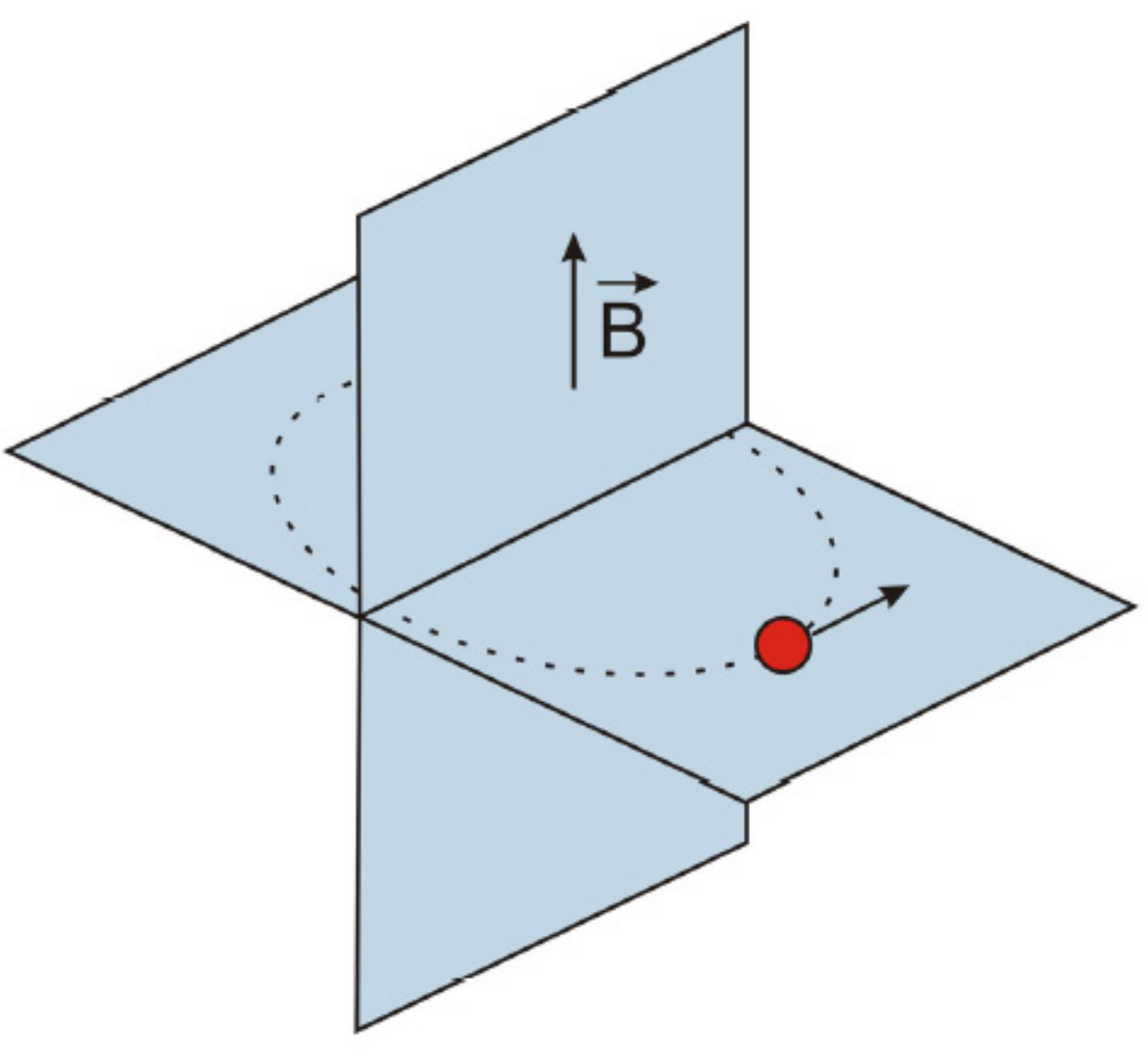}}}
\caption{\label{fig:Planes_unif_field} Schematics of the classical orbit of the electron
relative to the orientation of the uniform magnetic field.}
\end{figure}

\begin{figure}
\centering
\vspace*{- 3. cm}
{\scalebox{.45}{\includegraphics{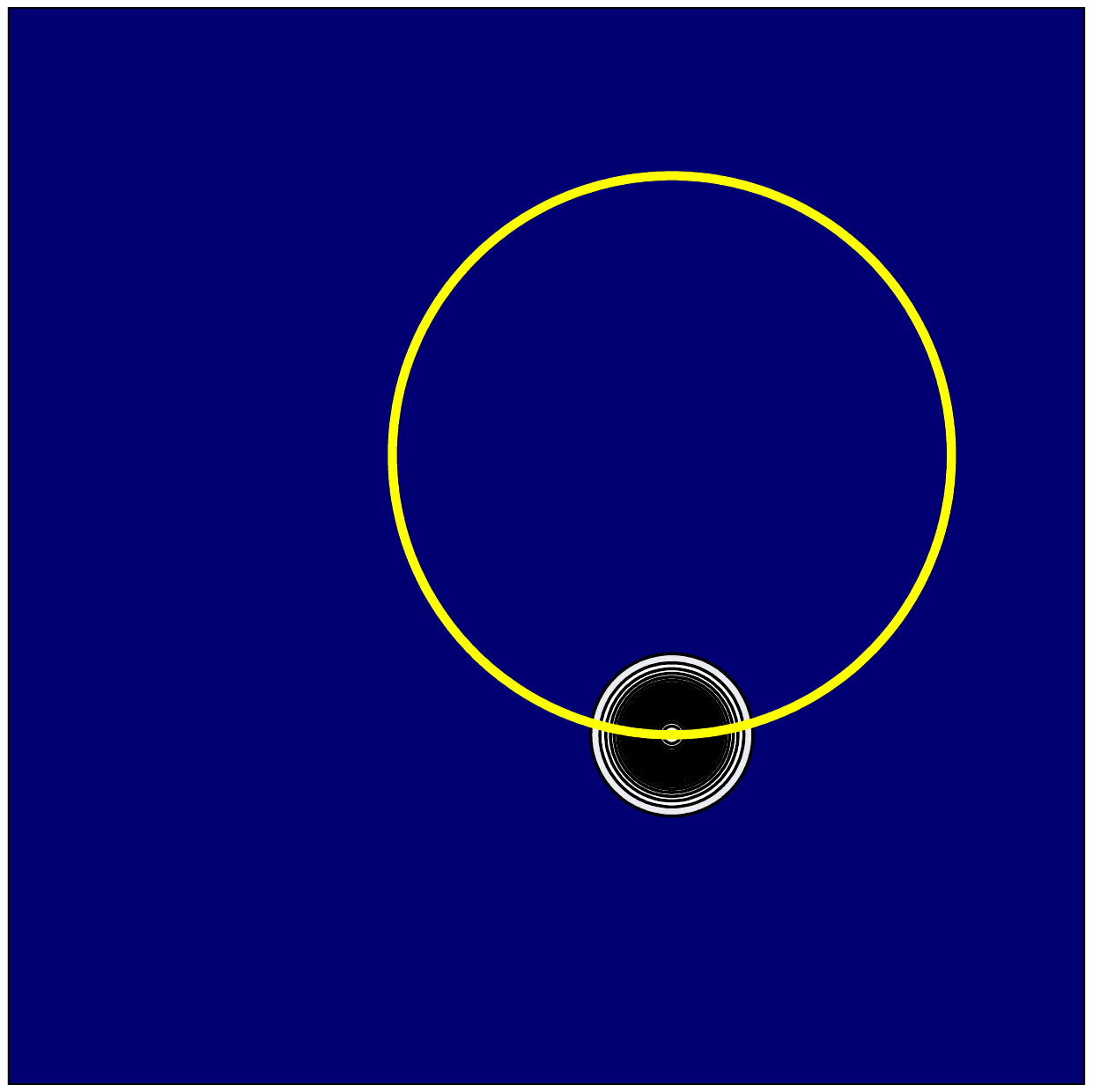}}}\hspace*{- 4.5 cm}
{\scalebox{.45}{\includegraphics{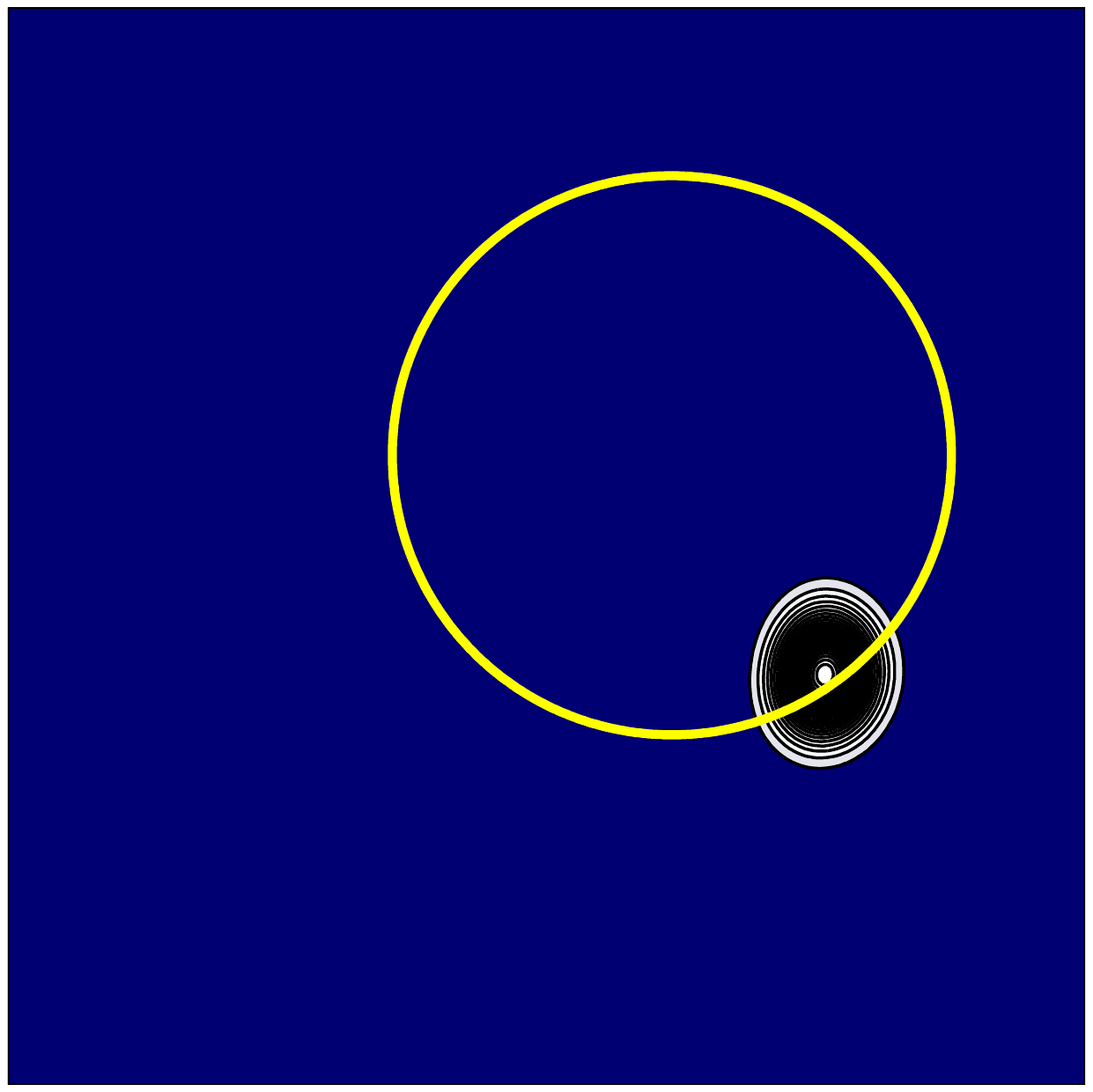}}}\vspace*{- 7.5 cm}
{\scalebox{.45}{\includegraphics{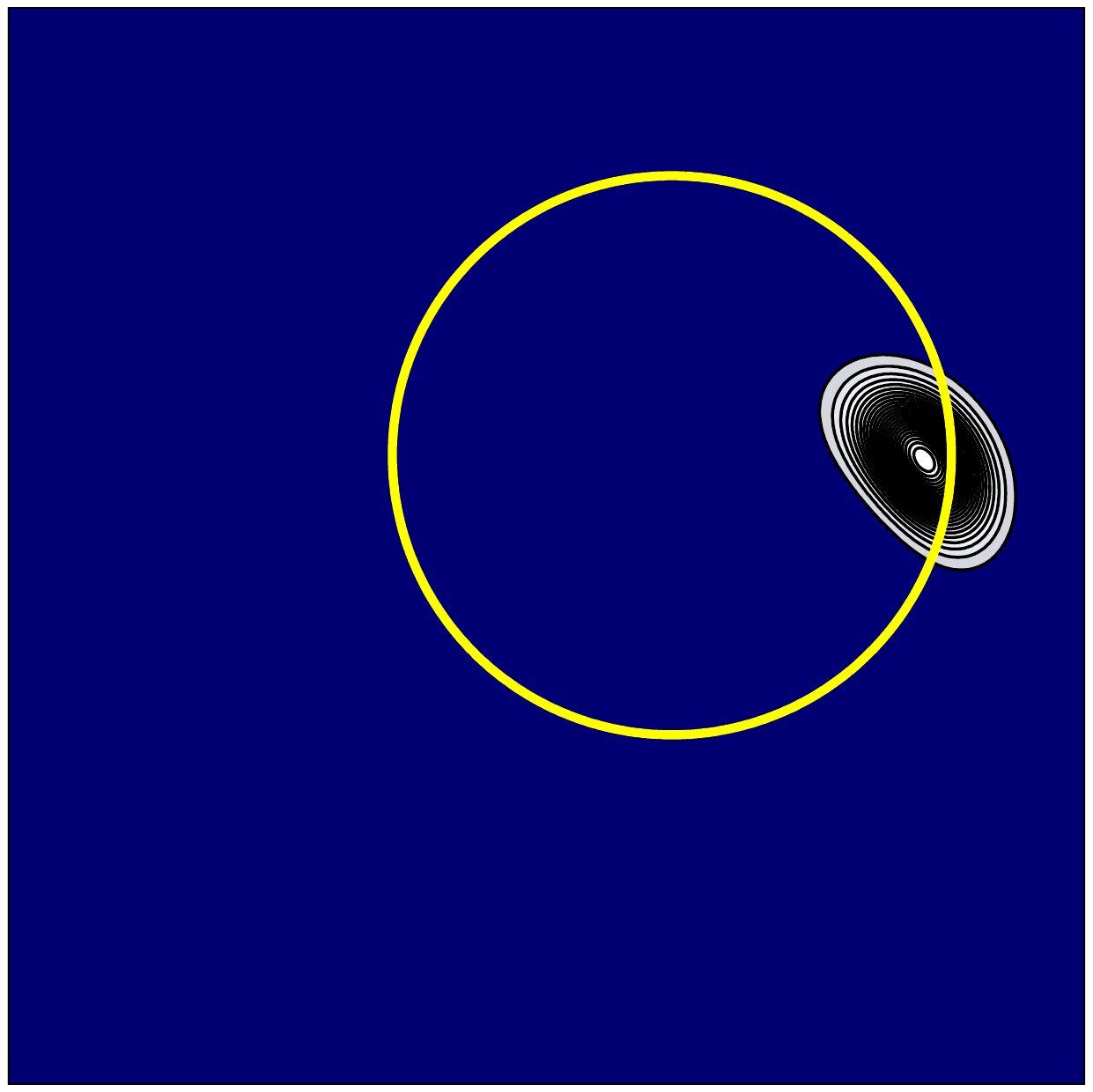}}}\hspace*{- 4.5cm}
{\scalebox{.45}{\includegraphics{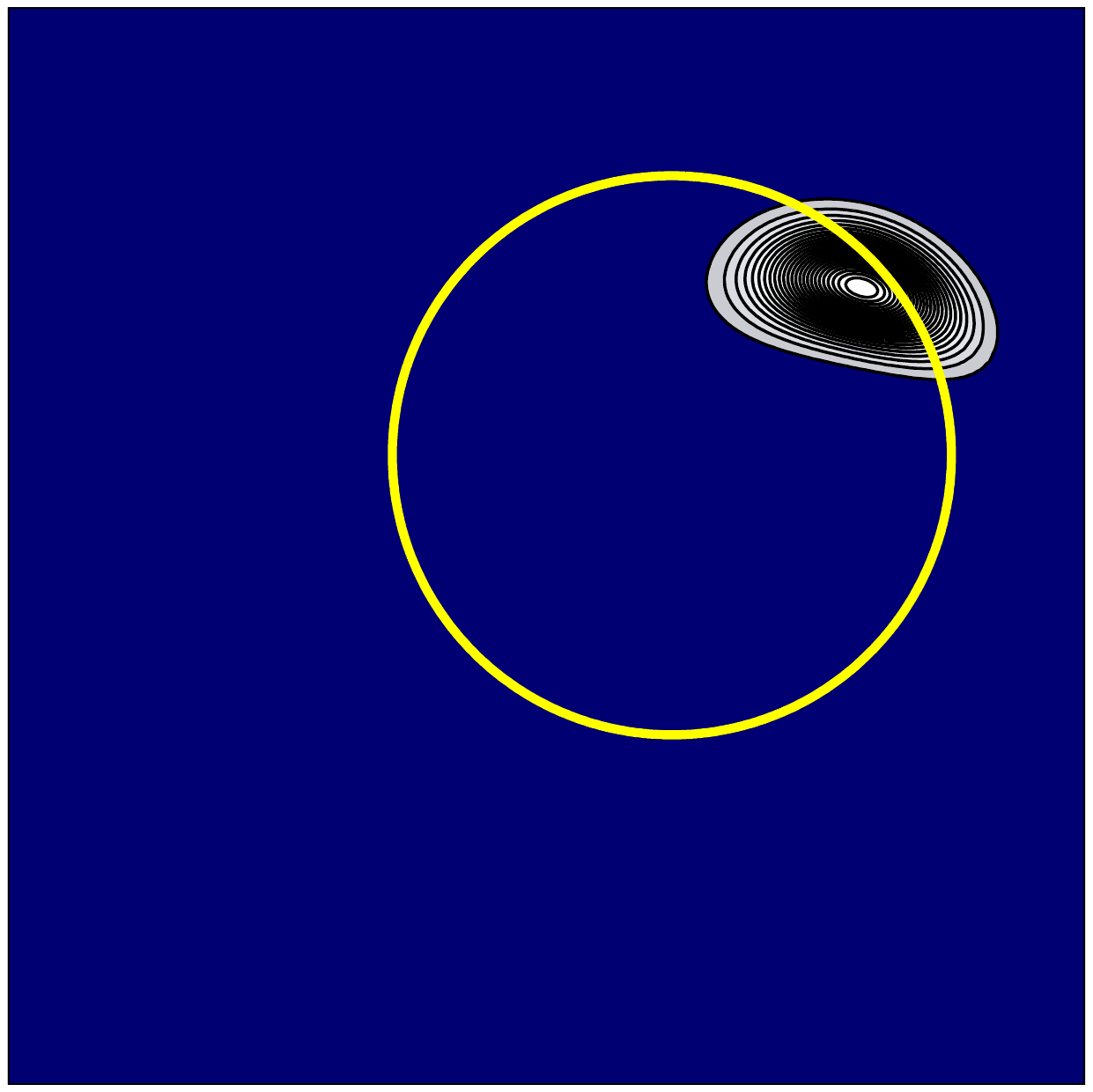}}}\vspace*{- 7.5 cm}
{\scalebox{.45}{\includegraphics{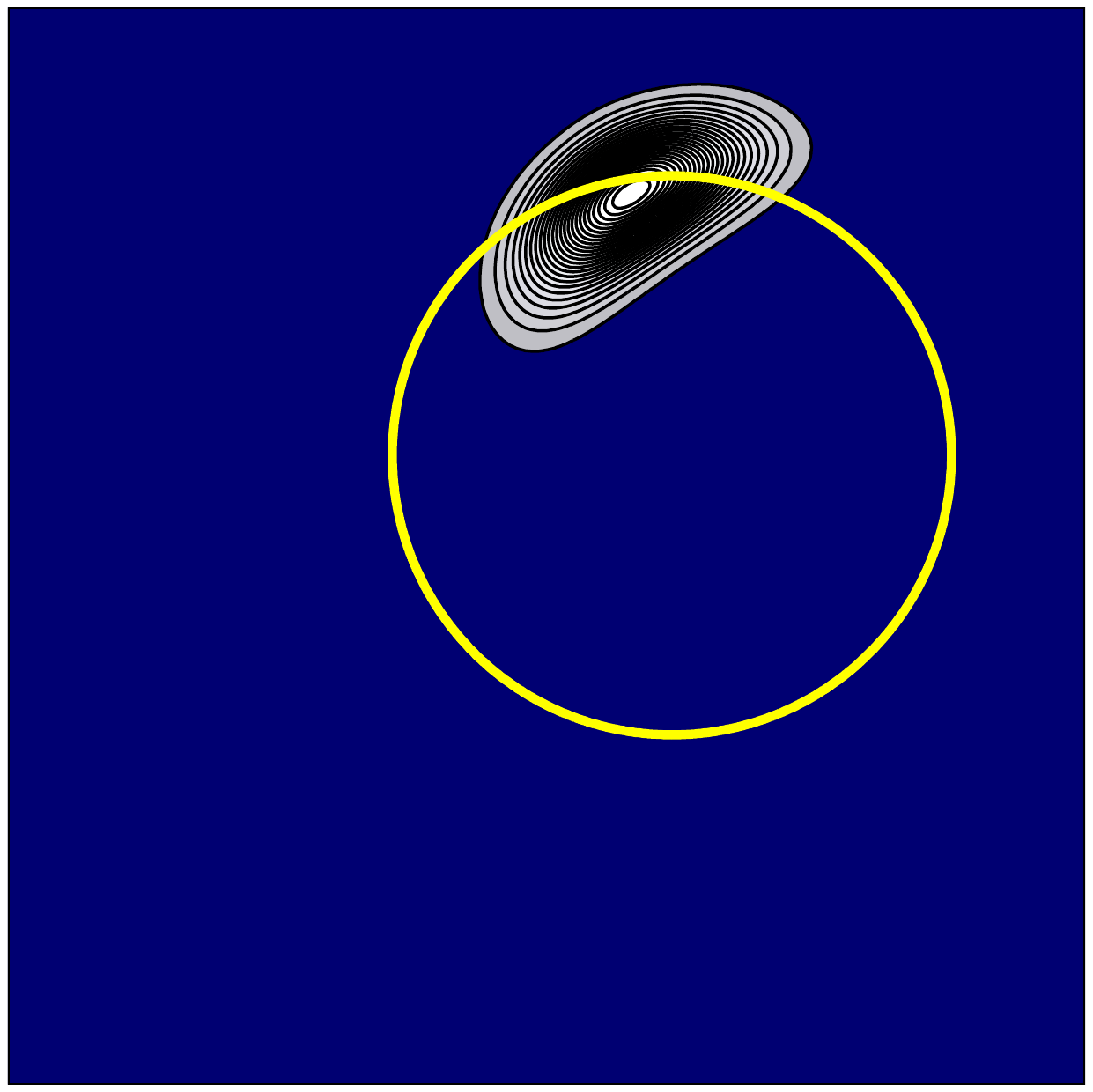}}}\hspace*{- 4.5 cm}
{\scalebox{.45}{\includegraphics{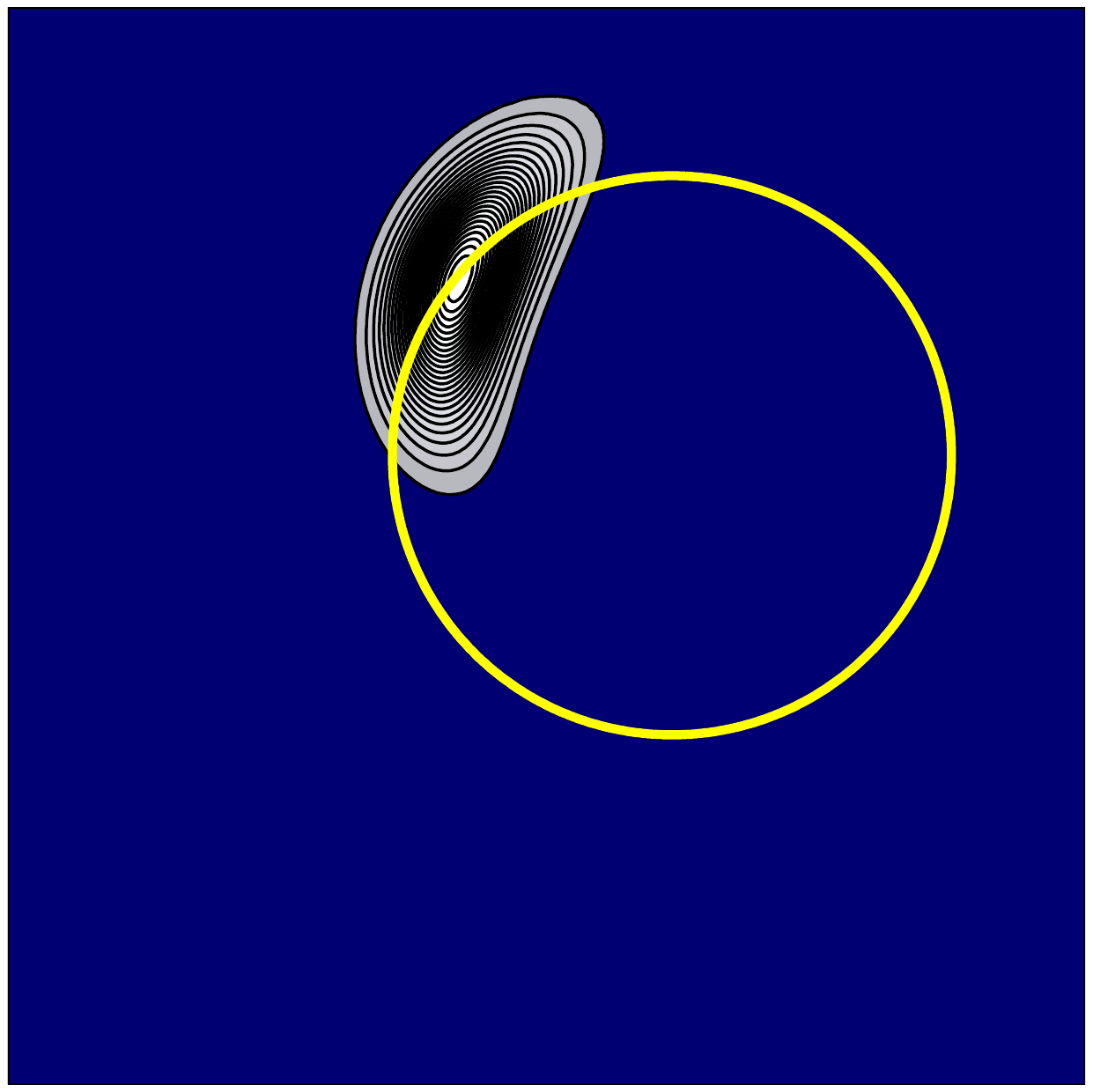}}}\vspace*{- 7.5 cm}
{\scalebox{.45}{\includegraphics{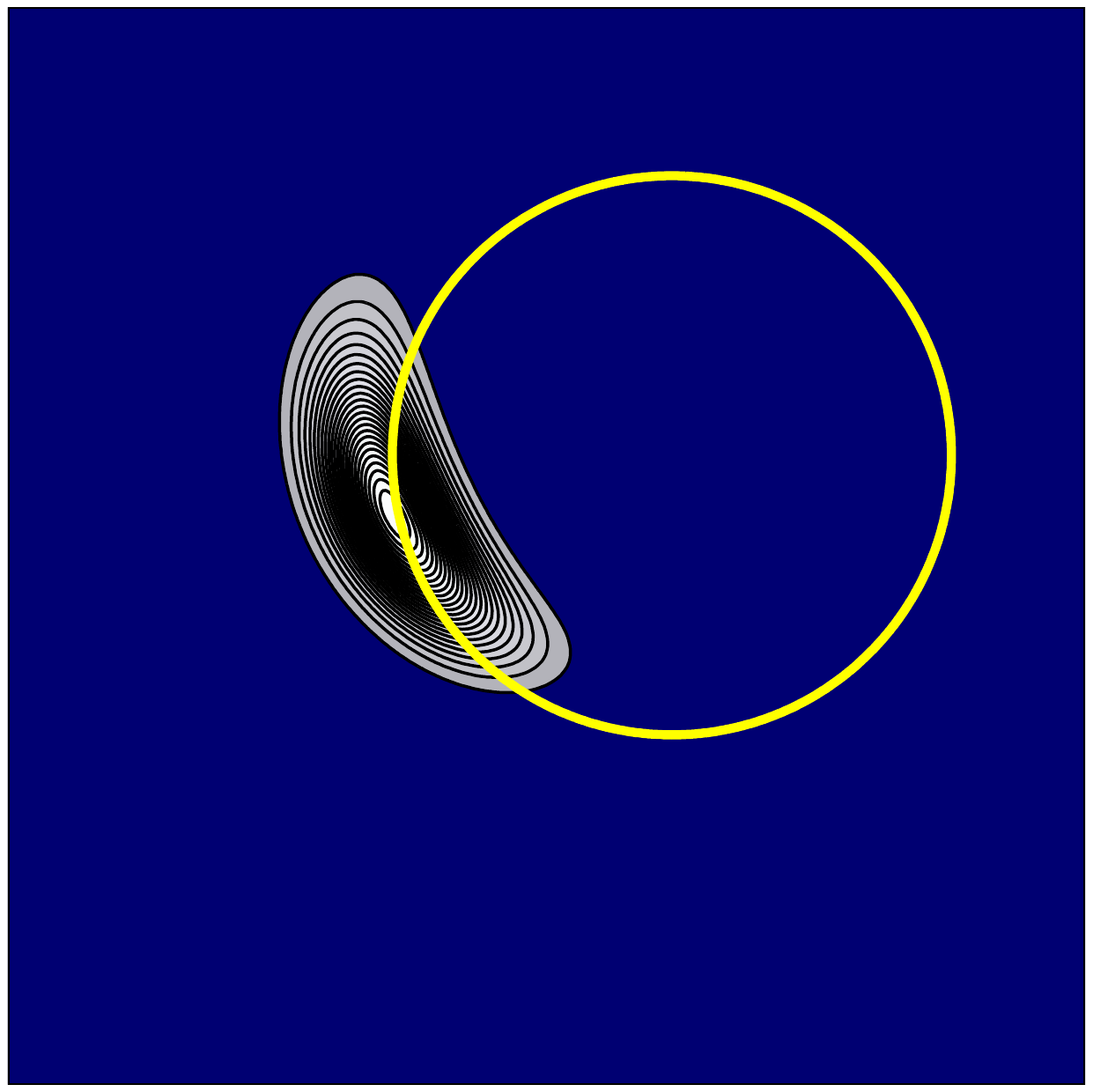}}}\hspace*{- 4.5 cm}
{\scalebox{.45}{\includegraphics{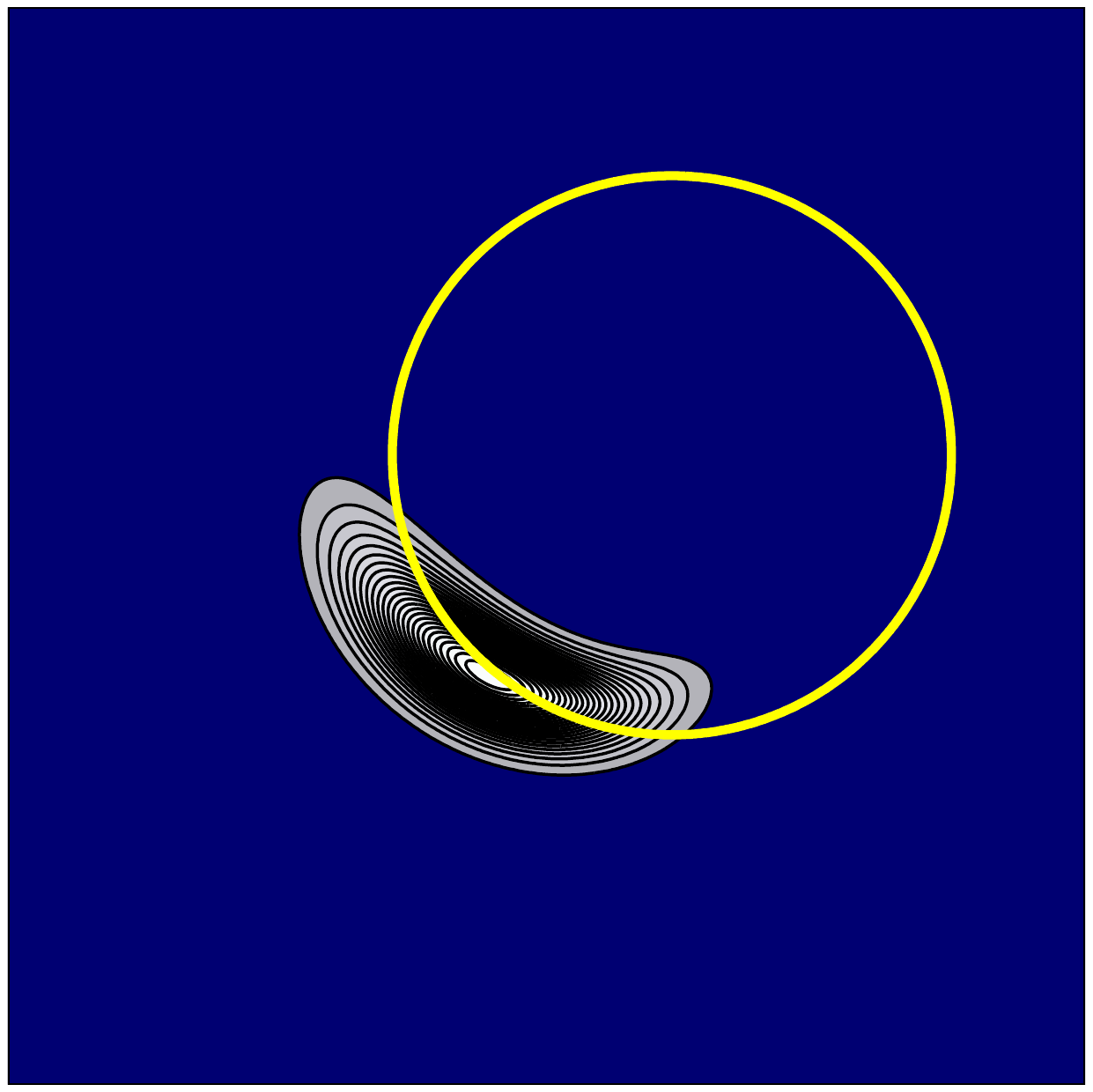}}}\vspace*{- 3. cm }
\caption{\label{fig:First_orbit_unif_field} The shapes and the positions of
the wave packet in the horizontal plane during the first rotation. The circle represents the classical orbit.
The animation can be accessed on-line \cite{Simi09b}.}
\end{figure}

\begin{figure}
\centering
\vspace*{- 4.5 cm}
{\scalebox{.5}{\includegraphics{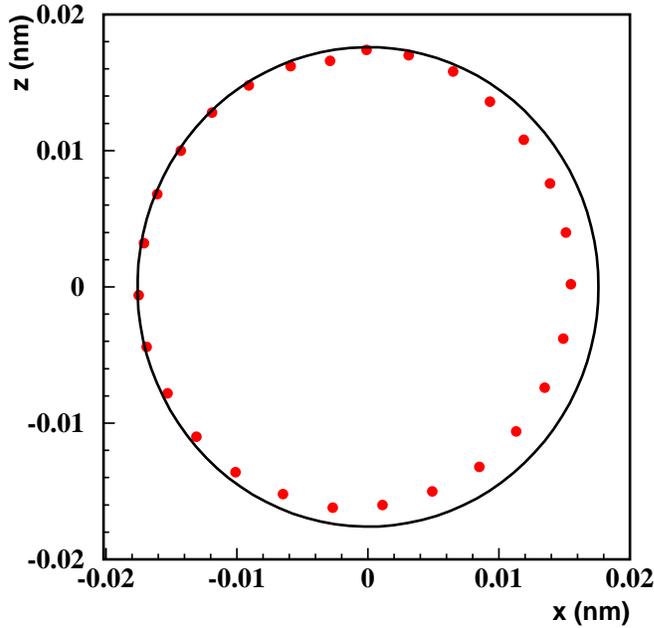}}}
\vspace*{- 1. cm}
\caption{\label{fig:First_orbit_position_cprob} Positions of the center of probability 
of the wave packet  in the horizontal plane during the first rotation. The circle represents the classical orbit.
The animation of the dynamics can be accessed on-line \cite{Simi09b}.}
\end{figure}

While the probability densities $|\Psi|^2$ were calculated for the entire computational volume and
at every time step, their values are shown here only
in the horizontal plane or plane of classical particle motion. 
The schematics of the wave packet or charged particle motion relative to the orientation of the magnetic field 
is shown in Figure \ref{fig:Planes_unif_field}. 
The position and the shape of the wave packet in the horizontal plane, as it moves along its first orbit,
is shown in Figure \ref{fig:First_orbit_unif_field}. The initial position of the wave packet was at the
center of the vector potential. Its initial momentum was $p_{1} = 0.53 \; MeV/c$, making the motion relativistic. 
In order to force a relativistic electron to complete a full circle in the available computational 
space, the magnitude of the magnetic field was $B_{0} = 10^{8} \; T$. The field of such 
a strength is associated with the fields at the surface of the neutron stars. The classical orbit of the electron in
this field is $r_{class}=p_{1}/(e B_{0})=1.76 \times 10^{-2}\; nm$. As shown in 
Figure \ref{fig:First_orbit_unif_field}, during the first rotation the wave packet 
generally follows the classical orbit, but disperses at the same time. The position of the center of probability 
of the wave packet relative to the classical orbit is shown in Figure \ref{fig:First_orbit_position_cprob}. 

\begin{figure}
\centering
{\scalebox{.45}{\includegraphics{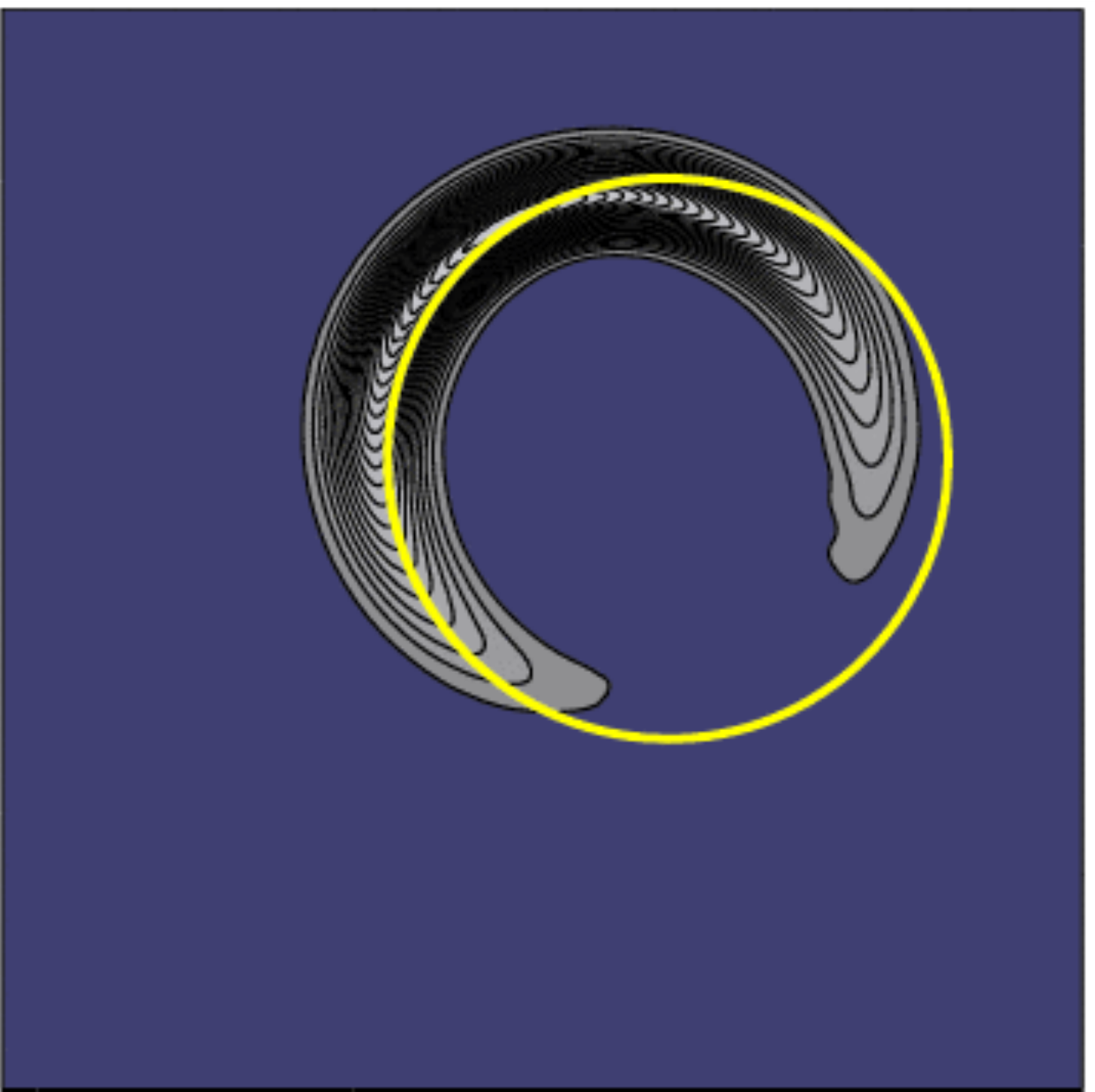}}}\hspace*{- 0.1 cm}
{\scalebox{.45}{\includegraphics{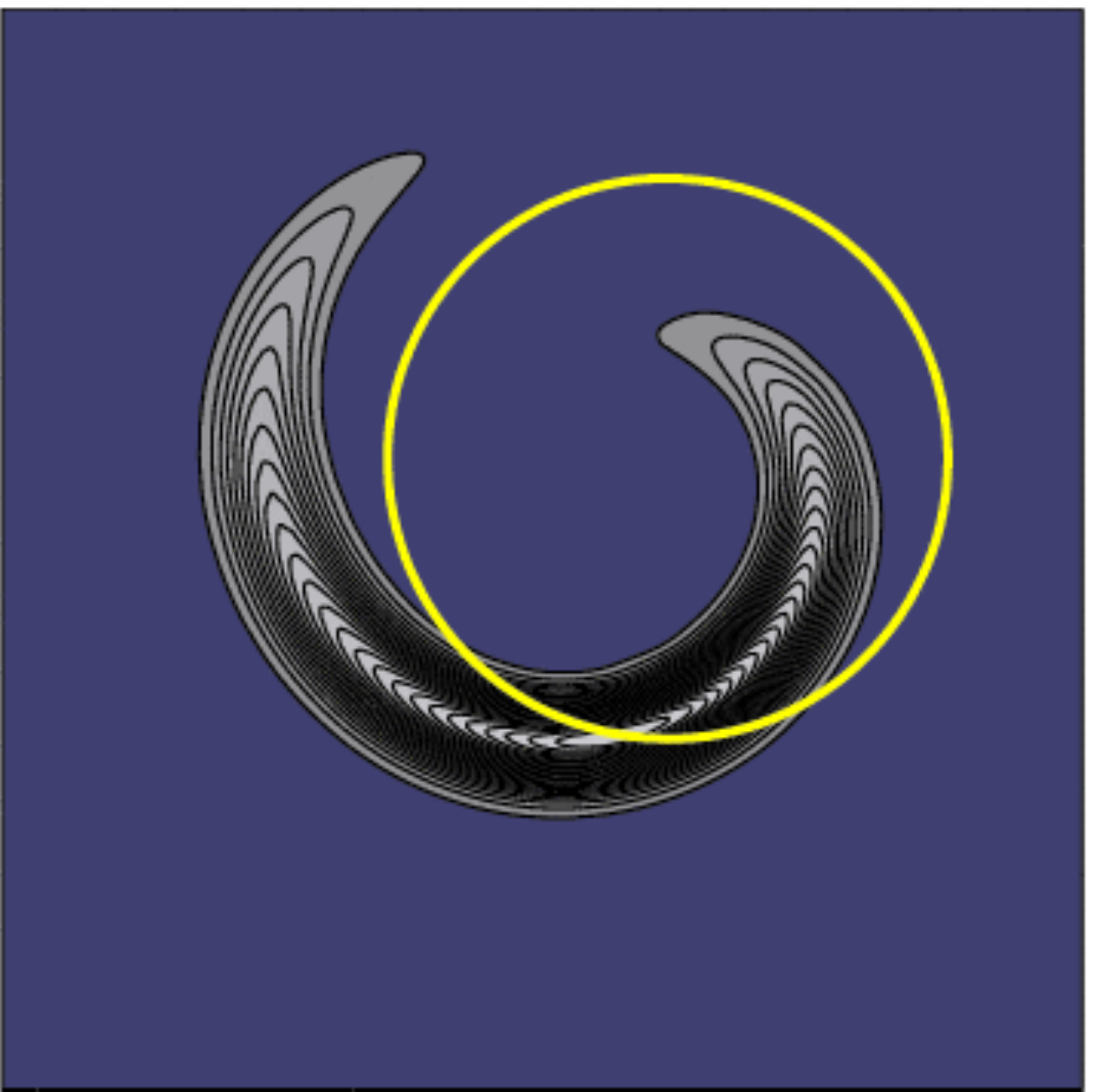}}}
{\scalebox{.45}{\includegraphics{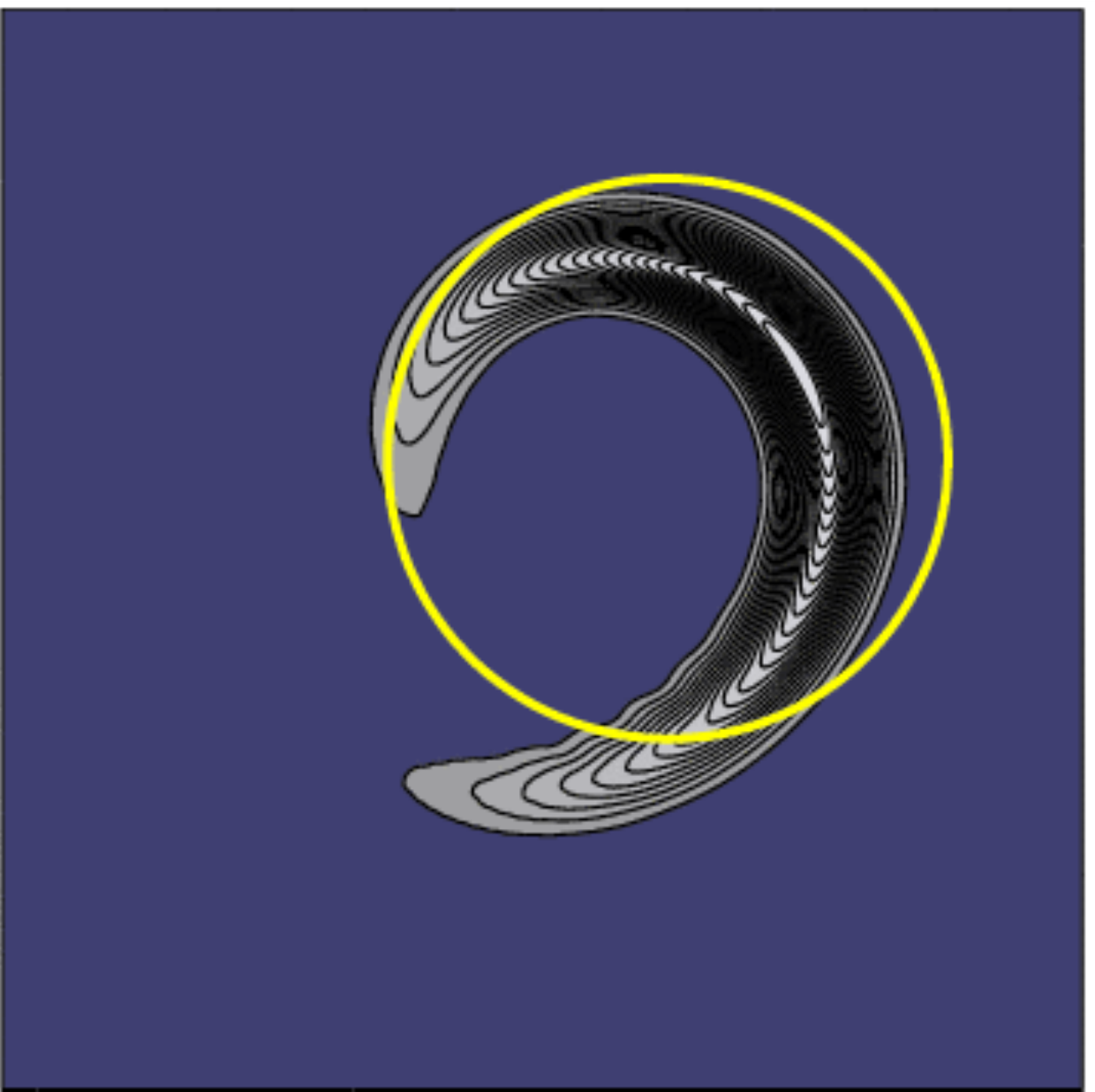}}}\hspace*{- 0.1 cm}
{\scalebox{.45}{\includegraphics{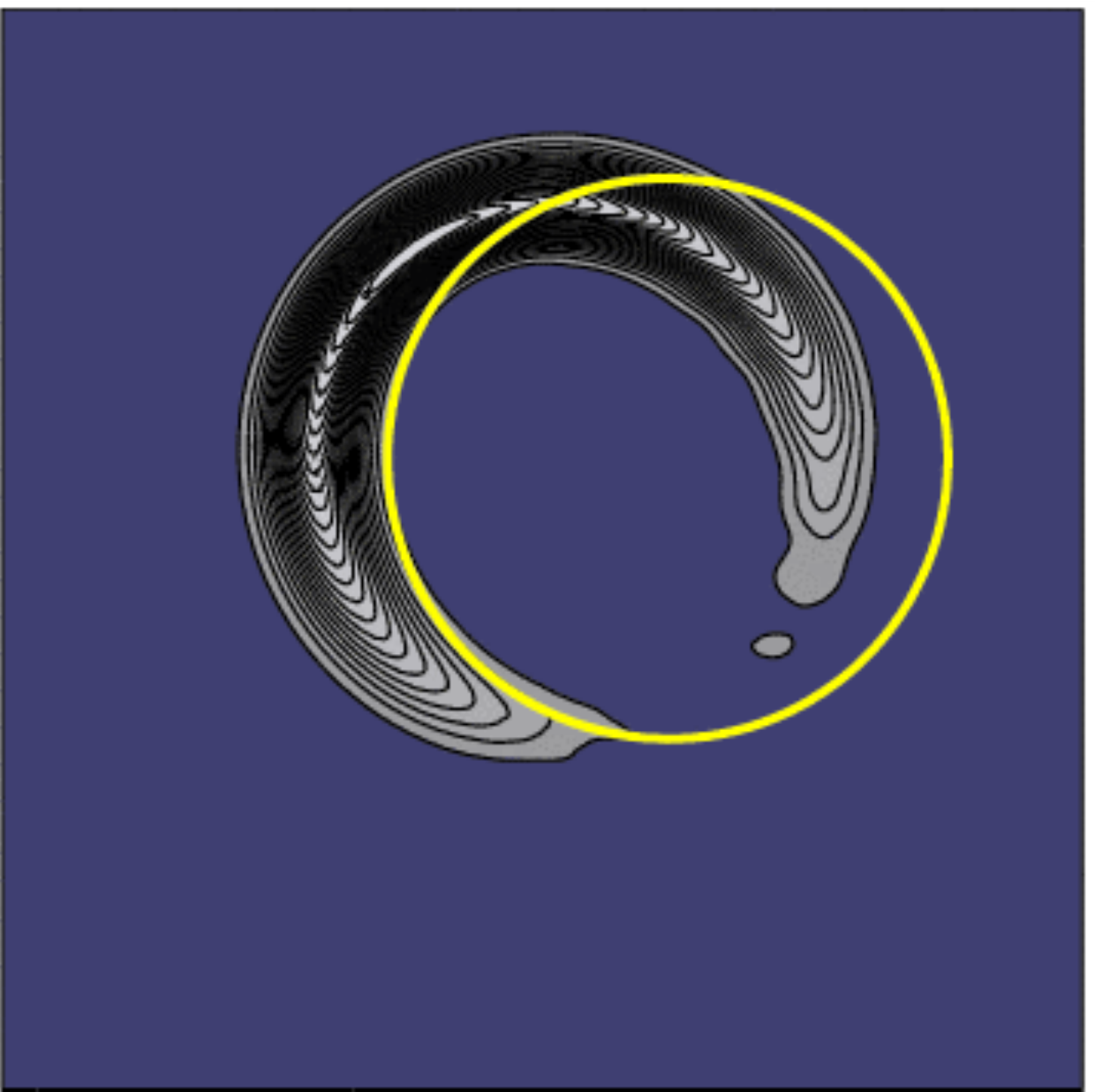}}}
\caption{\label{fig:Spiral_orbit_unif_field} Changes in the shape of the probability density $|\Psi|^2$
during one of the subsequent orbits of the wave packet. The circle represents the classical orbit.
The animation of the dynamics can be accessed on-line \cite{Simi09b}.}
\end{figure}

The complexity of the 
dynamics of the wave packet increased in the later stages of the motion. While the wave packet followed
the circular motion, the probability density $|\Psi|^2$ at some times assumed the spiral shape shown in 
Figure \ref{fig:Spiral_orbit_unif_field}, increased or reduced 
its length, changed its rotational motion, and translated  from one place to another. 
Overall, the characteristics of this dynamics were similar to the characteristics  described in 
Ref. \cite{Schl08} and \cite{Demi08}. While the figures show some of the complex shapes
of the probability density $|\Psi|^2$, the richness of the electron motion can be better appreciated through
the animation of the dynamics accessible on-line \cite{Simi09b}. 

Several additional tests of the FDTD method were performed using the dynamics of the 
electron in the uniform magnetic field. As expected, due to the normal spin and the magnetic field orientation,
reversal of the orientation of the 
magnetic field resulted in the reversal of the direction of the rotation of the wave packet, keeping the
same properties of the dynamics of motion. 

Of particular interest was testing the effects of the choice of gauge. 
The motion of the same wave packet was also studied for the 
uniform magnetic field oriented along the y-axis defined by the translationally invariant gauges 
\begin{equation}
\vec A={B_{0}} (z,0,0),
\label{Uni_mag_potent_x}
\end{equation}
or
\begin{equation}
\vec A={B_{0}} (0,0,-x).
\label{Uni_mag_potent_z}
\end{equation}

In both cases the wave packet persisted in a circular motion following the classical orbit. While
the dynamics of the wave packet behaved as expected, using translationally invariant gauges has 
an additional importance on validating the FDTD computation.
As seen in Eq. (\ref{Psi_1}), the gauge in Eq. (\ref{Uni_mag_potent_x}) couples to the $\Psi_{4}$ component
and the gauge in Eq. (\ref{Uni_mag_potent_z}) couples to the $\Psi_{3}$ component. Similarly, the cross coupling 
exists for other components of the wave function $\Psi$. As a result, the same dynamics should be obtained 
by different combinatorics of the components of the vector potential and the components of the wave function. 
This enabled for testing of possible inconsistencies in the FDTD updating equations. No  inconsistencies 
were found.

The dynamics of the wave packet motion in a uniform magnetic field was used here 
only as a validation 
of the FDTD method when a vector potential was applied in the Dirac equation. The same complexity 
of the quantum dynamics of motion as in previous publications \cite{Schl08,Demi08} was shown.
While used here only for computational validation of the FDTD method, this dynamics could
be studied as a separate problem in more detail in the future.
    
Finally, the complexity of the quantum dynamics
of the wave packet in a uniform magnetic field studied here for three choices of gauge
could be fully appreciated only by downloading the related animations \cite {Simi09b}. 
  
\subsection{Motion of a wave packet between two infinite lines of electric dipoles}

The motion of a wave packet in a uniform magnetic field is used to validate 
the FDTD method when a vector potential $\vec A(x)$ is applied in the Dirac equation.
The motion of a wave packet  between two infinite lines of electric dipoles is used to
validate the dynamics
of a wave packet under influence of a position-dependent electromagnetic scalar potential $A_{0}(x)$, 
and to validate 
the geometry associated with the Aharonov-Bohm effect studied in the next section. 
In the case of a wave packet moving between two infinite lines of electric dipoles,
the dynamics obtained from 
the solution of the Dirac equation can be compared 
to the dynamics of a charged particle influenced by a classical electromagnetic force. The
classical dynamics, resulting from the interaction of a charged particle and two infinite lines 
of electric dipoles, is obtained following Boyer \cite{Boy02}.  

The scalar potential $A_{0}$ of two infinite lines of 
electric dipoles parallel to the y-axis and separated by a distance $2a$, with dipoles 
oriented in the x-direction, parallel or anti-parallel to the direction of the wave packet motion, is

\begin{equation}
A_{0}= \pm {\wp \over 2\pi\epsilon_{0}} \left ( {{x} \over {x^{2} + (z-a)^{2}}}+ {{x} \over{x^{2} + (z+a)^{2}}}
\right ),
\label{a0_lines_of_dipoles}
\end{equation}
where $\wp$ is the line density of dipoles and $\epsilon_{0}$ is the permittivity constant. 
With the choice of  $z=0$, the classical motion of a charged particle along the x-axis, 
under influence of this potential, can be calculated by solving the equation

\begin{equation}
{d p_{x} \over dt} =  - q {\partial A_{0} \over \partial x}
=\mp {q \wp \over \pi\epsilon_{0}} {1 \over {x^{2} + a^{2}}} 
\left (1-{2x^{2} \over {x^{2} + a^{2}}}\right )
\label{classical_force}
\end{equation}

\begin{figure}
\centering
\vspace*{- 0.5 cm}
\hspace*{- 1.2 cm}
{\scalebox{.5}{\includegraphics{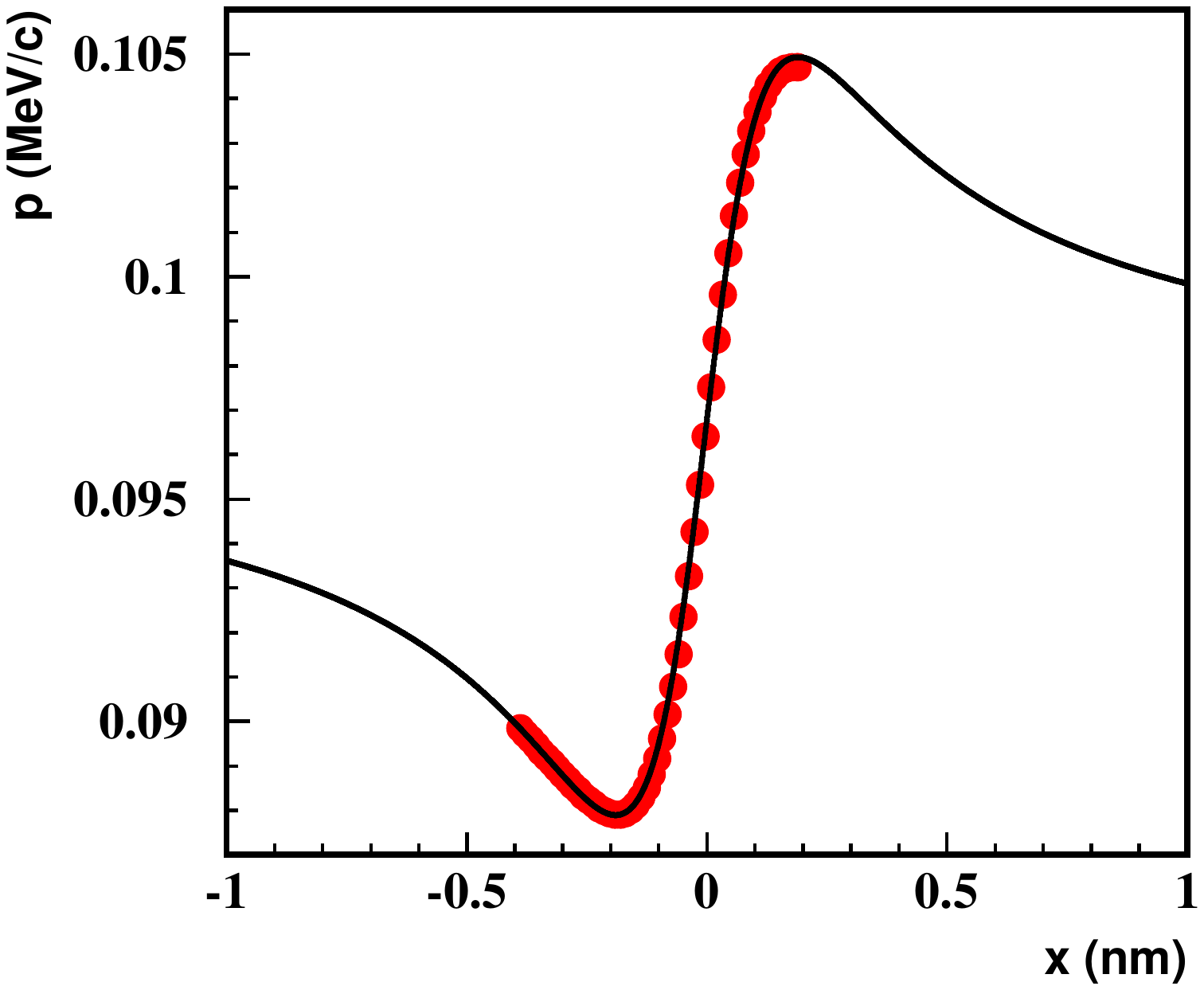}}}\hspace*{- 2.75 cm}
{\scalebox{.5}{\includegraphics{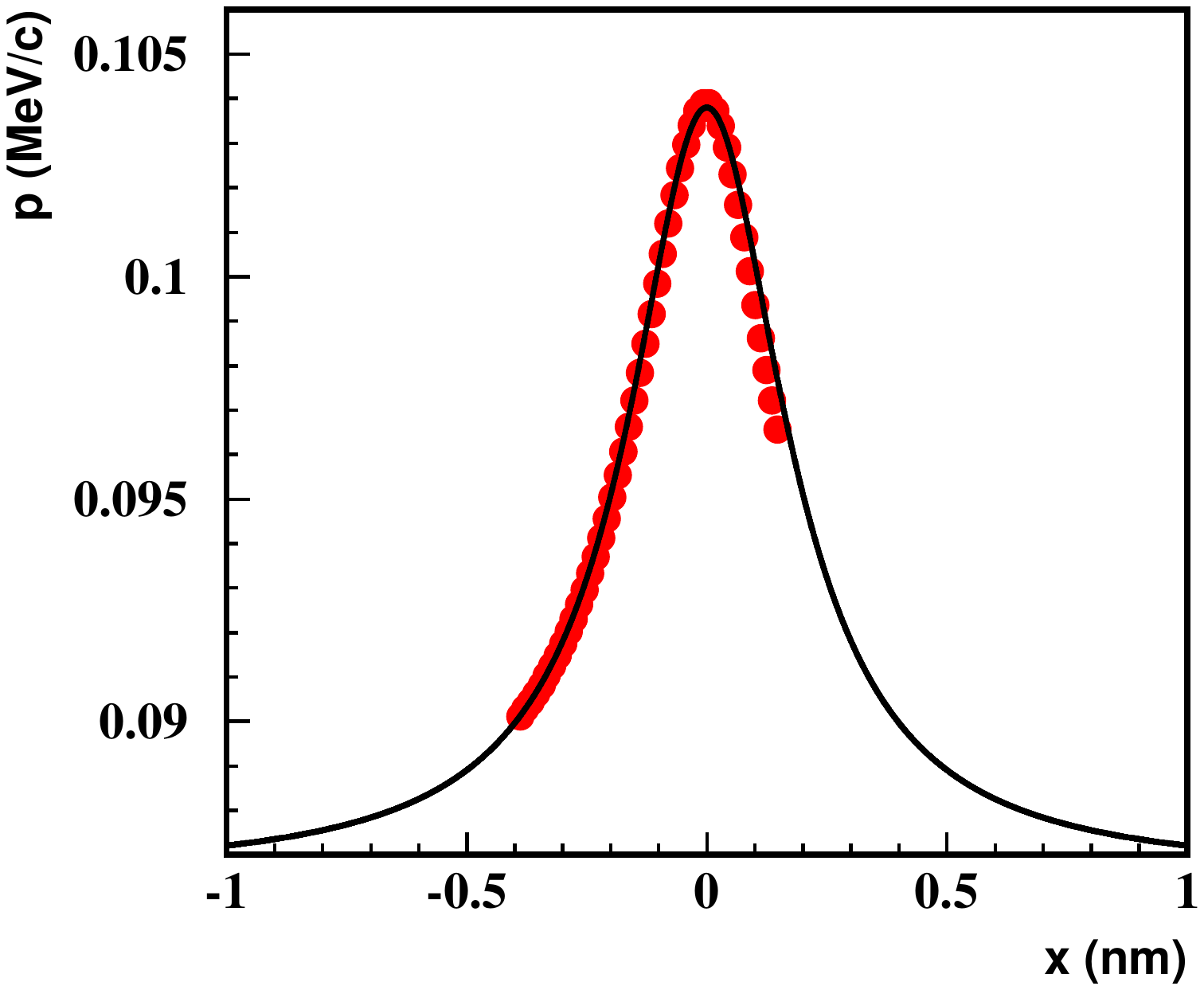}}}
\vspace*{- 5.5 cm}
\caption{\label{fig:lines_of_dipoles_test} On the left side, the solution of the Dirac equation
with the scalar potential $A_{0}$ described by Eq. (\ref{a0_lines_of_dipoles}) is compared to 
the solution of  Eq. (\ref{classical_force}). The right side compares the solution of the Dirac equation
with the scalar potential described by Eq. (\ref{a0_lines_of_dipoles_normal}) to 
the solution of  Eq. (\ref{classical_force_normal}).
The red circles rapresent the solutions of the Dirac equation and the solid lines the classical solutions.
In both cases the momentum of the charged particle was $p=0.09 \; MeV/c$ at the position $x=-0.4 \; nm$.
The distance between the  lines of electric dipoles was $2a=0.38 \; nm$ and the 
line density of dipoles  was $\wp = 1.7 \; 10^{-17} \; Cm/m$.
In the coordinate system used in this paper two infinite lines of 
electric dipoles were parallel to the y-axis and the wave packet moved along the x-axis.}
\end{figure}

The comparison of the quantum-mechanical dynamics obtained as a solution of the Dirac equation
with the scalar potential described by Eq. (\ref{a0_lines_of_dipoles})
and the classical dynamics obtained as a solution of  Eq. (\ref{classical_force}) is shown on the left side in 
Figure \ref{fig:lines_of_dipoles_test}. The relativistic effects in the classical dynamics 
were reduced to  less than $2 \%$ by choosing the initial momentum $p=0.096 \; MeV/c$. 
The lines of electric dipoles were separated by a distance $2a=0.38 \; nm$ and the 
line density of dipoles  $\wp$ was $1.7 \; 10^{-17} \; Cm/m$. Figure \ref{fig:lines_of_dipoles_test}
shows agreement between the quantum-mechanical and classical dynamics over the 
distance for which the solution of the Dirac equation was computed.

Under the same conditions, the computation was repeated for dipoles 
oriented in the z-direction, anti-parallel to each other, and perpendicular to the direction 
of the wave packet motion. In this case the 
scalar potential $A_{0}$ is  
\begin{equation}
A_{0}= \pm {\wp \over 2\pi\epsilon_{0}} \left ( {{z-a} \over {x^{2} + (z-a)^{2}}}- {{z+a} \over{x^{2} + (z+a)^{2}}}
\right ),
\label{a0_lines_of_dipoles_normal}
\end{equation}
and the equation of motion, for $z=0$, is 
\begin{equation}
{d p_{x} \over dt} =\mp {q \wp \over \pi\epsilon_{0}} {2ax \over {(x^{2} + a^{2})^{2}}}.
\label{classical_force_normal}
\end{equation}
As shown on the right side in Figure \ref{fig:lines_of_dipoles_test}, the solution of the Dirac equation
with the scalar potential described by Eq. (\ref{a0_lines_of_dipoles_normal}) and the solution of  Eq. (\ref{classical_force_normal})
agree again over the range of computation.

\section{Wave packet dynamics in a vector potential created by two infinite solenoids}

The goal of this work is to study the dynamics 
of an electron wave packet under the influence of a vector potential associated 
with the Aharonov-Bohm effect \cite{AB59}. Particularly, in this paper we present 
the dynamics of a wave packet obtained from the solution of the Dirac equation 
with a vector potential created by two infinite solenoids. 

\begin{figure}
\centering
\vspace*{- 0.5 cm}
{\scalebox{.45}{\includegraphics{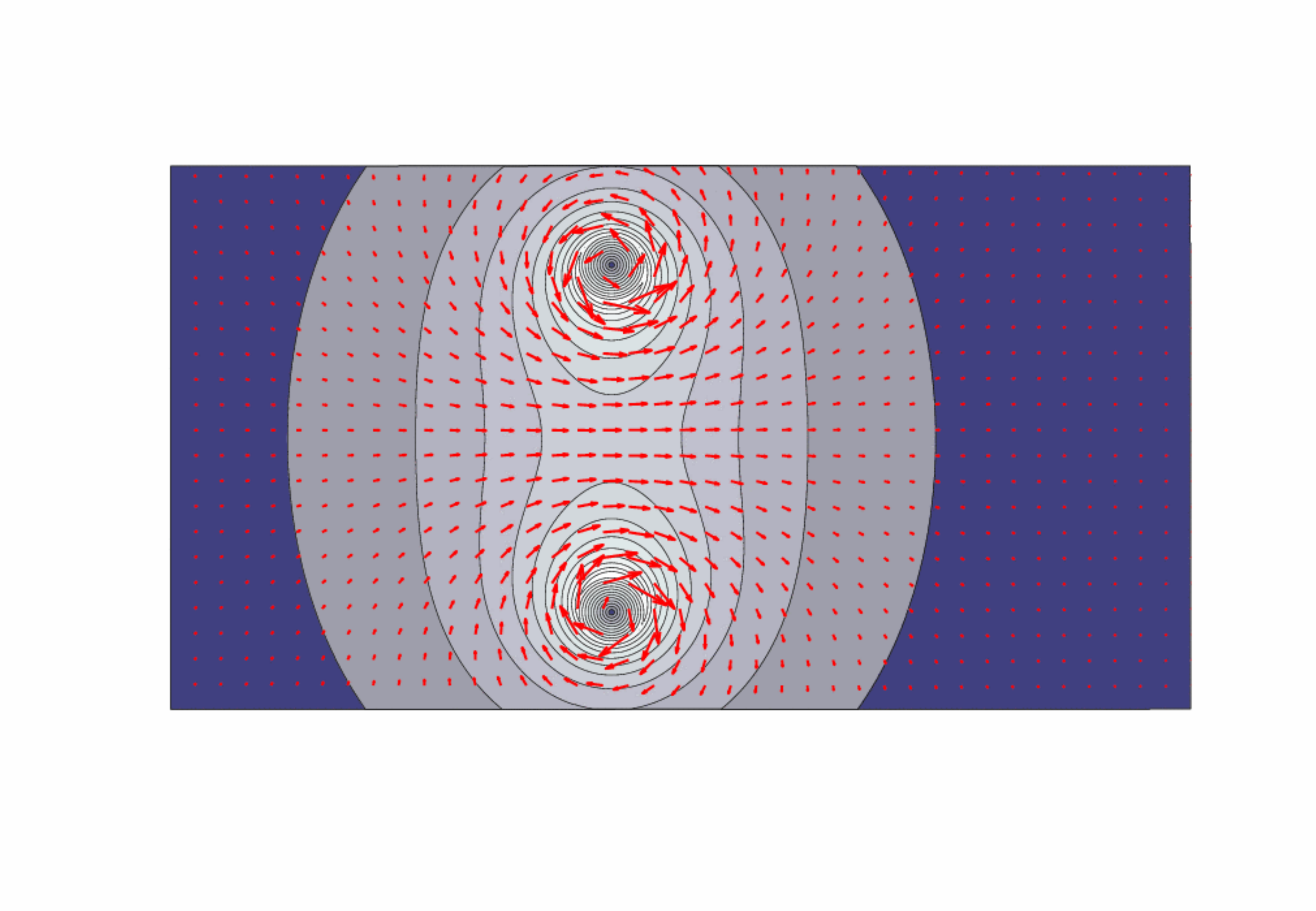}}}
\vspace*{- 2.8 cm}
\caption{\label{fig:plot_of_vector_potential} The shape of the vector potential of two parallel infinite solenoids
with oppositely directed magnetic fluxes of the same magnitude. The potential is shown in the plane
normal to the orientation of the solenoids. In the coordinate system used in this paper the magnetic field of the 
solenoids is oriented along the y-axis, the vector potential lays in the x-z plane, 
while the wave packet moves along the x-axis.}
\end{figure}

The vector potential of a single infinite solenoid oriented along the y-axis can be written as

\begin{equation}
{\vec A =\left\{ \begin{array} {c} {\Phi \over {2 \pi R_{0}^{2}}} (z,0,-x) \;\;\;\;\; \mbox{ for $r \leq R_{0}$} \\ 
{\Phi \over {2 \pi r^{2}}}(z,0,-x) \;\;\;\;\; \mbox{ for $r > R_{0}$}
 \end{array} \right. },
\label{Vect_pot_inf_solenoid}
\end{equation}
where $\Phi=B_{0} \pi R_{0}^{2}$ is the magnetic flux inside the solenoid, $B_{0}$ defines the 
strength of the magnetic field, $R_{0}$ is the radius of the solenoid,  and
$r=\sqrt{x^{2} + z^{2}}$ is the distance from the center of the solenoid in the x-z plane. Outside two 
parallel infinite solenoids separated by a distance $2a$ and with opposite magnetic field 
orientation, the vector potential is then

\begin{equation}
\! \! \! \! \! \! \! \! \! \! \! \! \! \! \! \! \! \! \vec A= {\Phi \over {2 \pi}} \left ( {{z-a} \over {x^{2} + (z-a)^{2}}}- {{z+a} \over{x^{2} + (z+a)^{2}}}
,0,{{x} \over{x^{2} + (z+a)^{2}}}-{{x} \over{x^{2} + (z-a)^{2}}}\right ) \!  .
\label{Vect_pot_two_inf_solenoid}
\end{equation}
An example of the shape of this potential is shown in Figure \ref{fig:plot_of_vector_potential}.
Outside the solenoids, the associated magnetic fields $\vec B = \vec \nabla \times \vec A$ 
of the vector potentials described by Eqs. (\ref{Vect_pot_inf_solenoid}) 
and (\ref{Vect_pot_two_inf_solenoid}) are zero. 

The time-dependent Dirac equation is now solved for the scalar component 
of the four-potential $A_{0}=0$ and the vector component $\vec A$ defined by  
Eq. (\ref{Vect_pot_two_inf_solenoid}). With this choice  we  can study the dynamics of the 
wave packet in the region where the electric and the magnetic fields of the unperturbed
solenoids are zero. 
 
The dynamics of the electron motion between two infinite and parallel solenoids is essentially the same as the
electron dynamics in the case of Aharonov-Bohm effect. In both cases the electron moves in a region 
where the unperturbed solenoids produce no Lorentz force on the electron. Because of the symmetry, 
however, the dynamics in the case of two infinite solenoids consists 
only of the motion along one straight line between the solenoids. This avoids the complications of 
the  Aharonov-Bohm dynamics where the trajectories on opposite sides of a single solenoid are compared.  
Because the properties of the dynamics are not studied through an interference, 
in the case of two solenoids the changes
in the dynamics of the electron wave packet exclude the contribution of the nonlocal properties of the electron 
wave function dependent on the topology of the space. Another advantage of using two solenoids 
and localized wave packet is the possibility to separate the solenoids far enough that no parts of 
the wave packet penetrate the non-zero field region inside the solenoids, excluding this contribution to the dynamics 
as well. If necessary, the solenoids could be shielded by a potential barrier of a supercritical potential 
additionally preventing any penetration of the wave packet into non-zero field region.

\begin{figure}
\centering
\vspace*{- 9. cm}
{\scalebox{.55}{\includegraphics{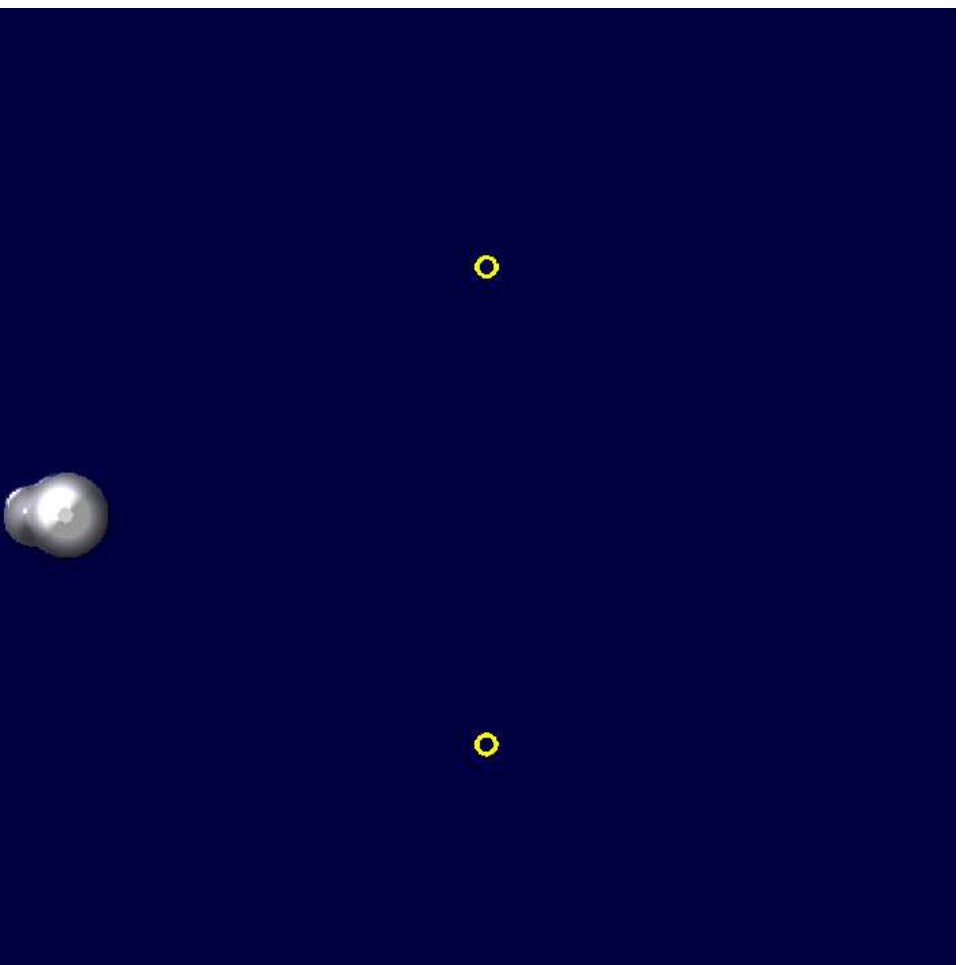}}}\hspace*{- 6.5 cm}
{\scalebox{.55}{\includegraphics{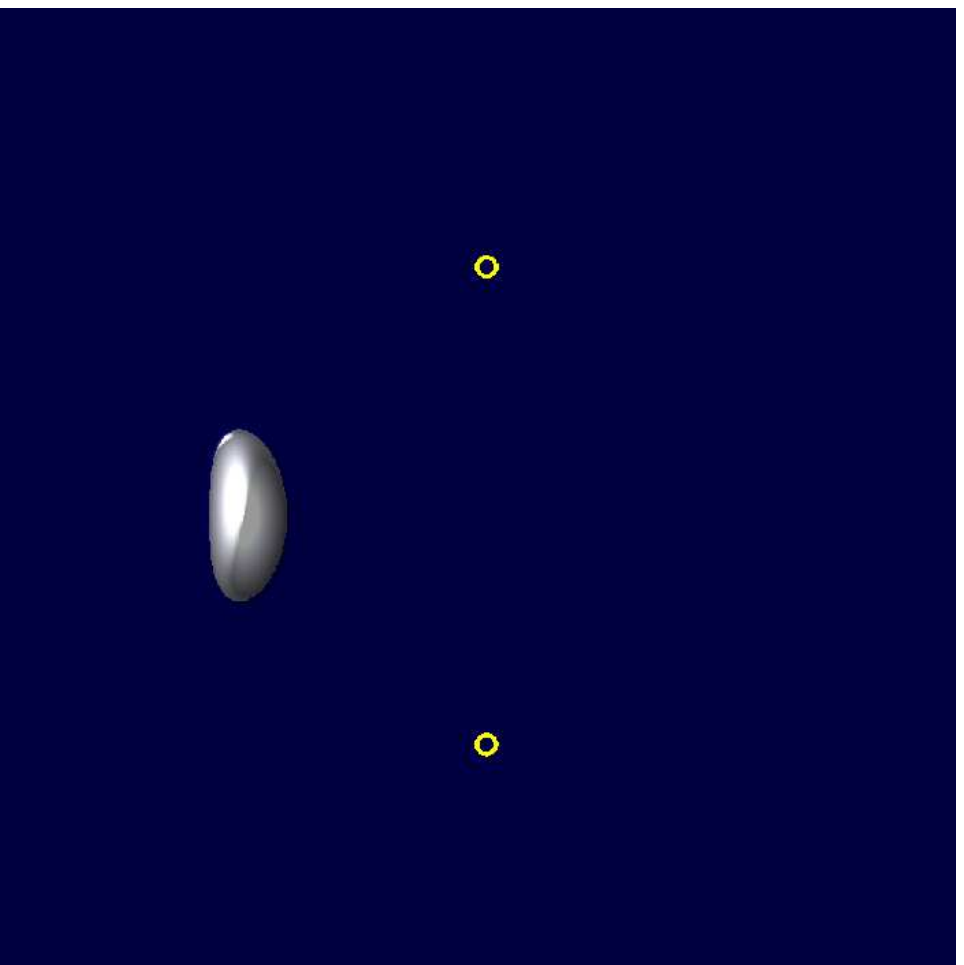}}}\hspace*{- 6.5 cm}
{\scalebox{.55}{\includegraphics{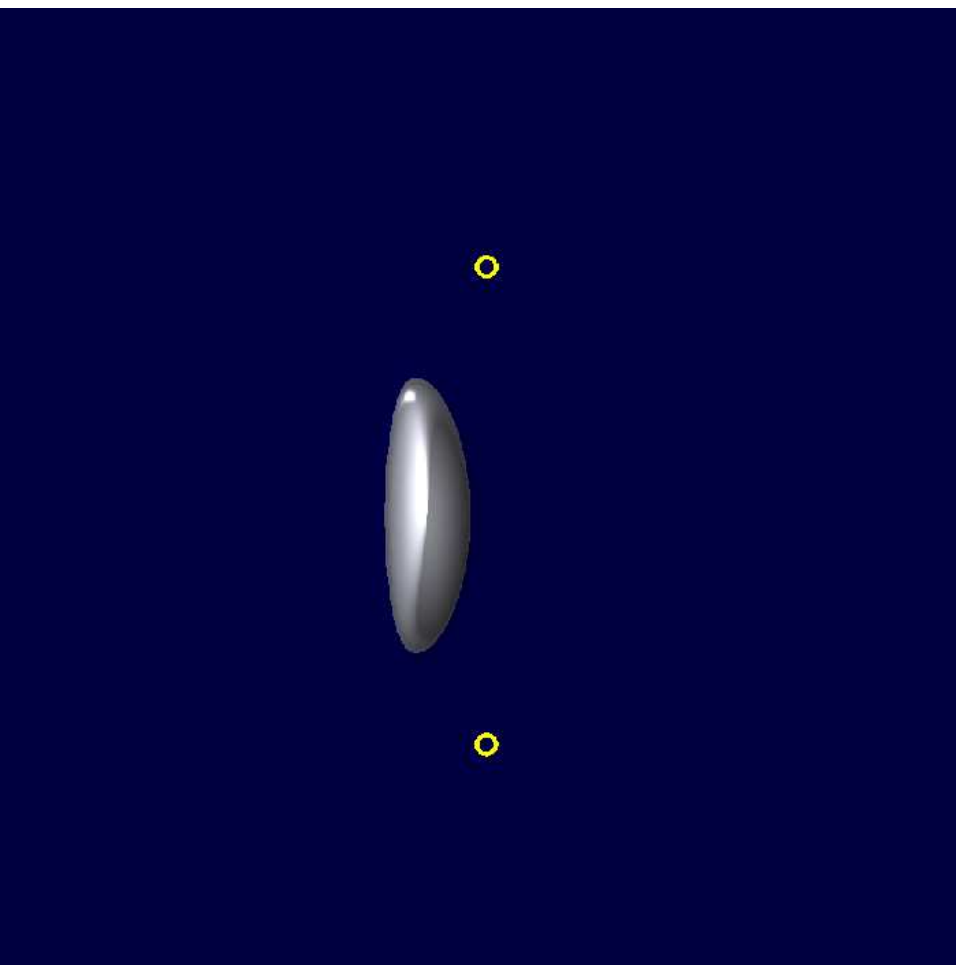}}}\vspace*{- 9.75 cm}                   
{\scalebox{.55}{\includegraphics{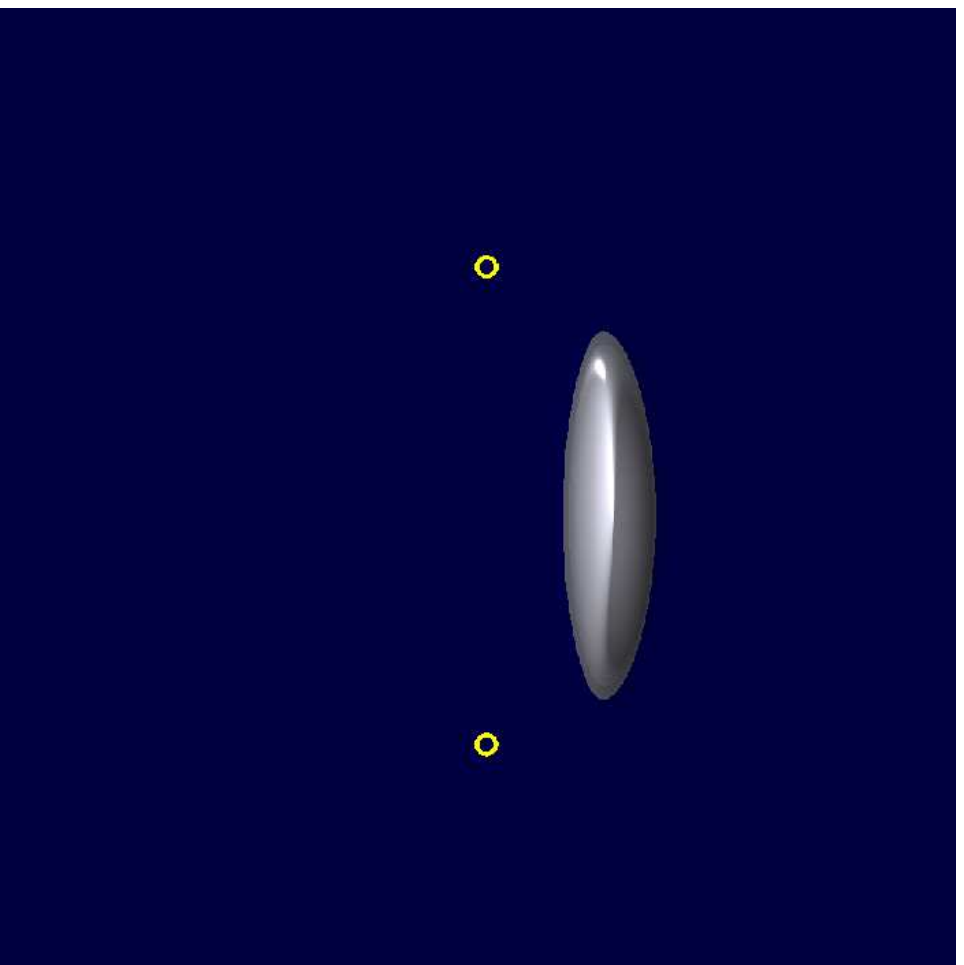}}}\hspace*{- 6.5 cm}
{\scalebox{.55}{\includegraphics{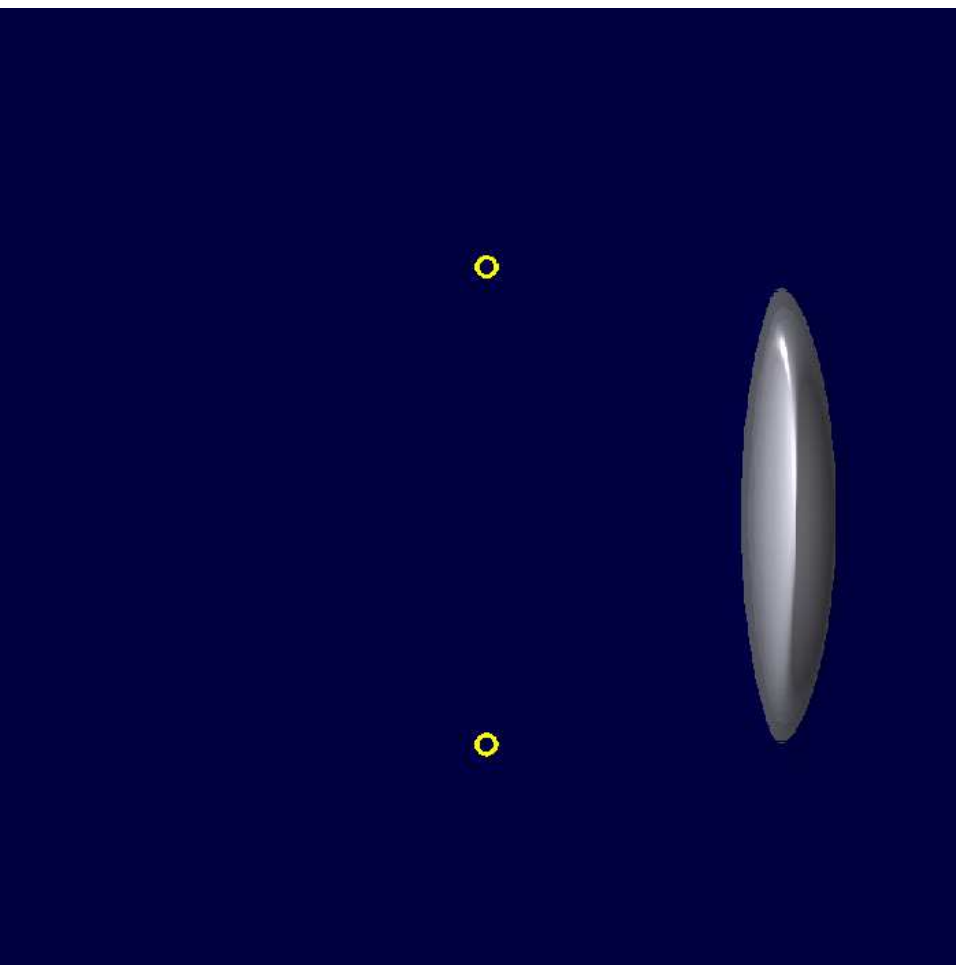}}}\hspace*{- 6.5 cm}
{\scalebox{.55}{\includegraphics{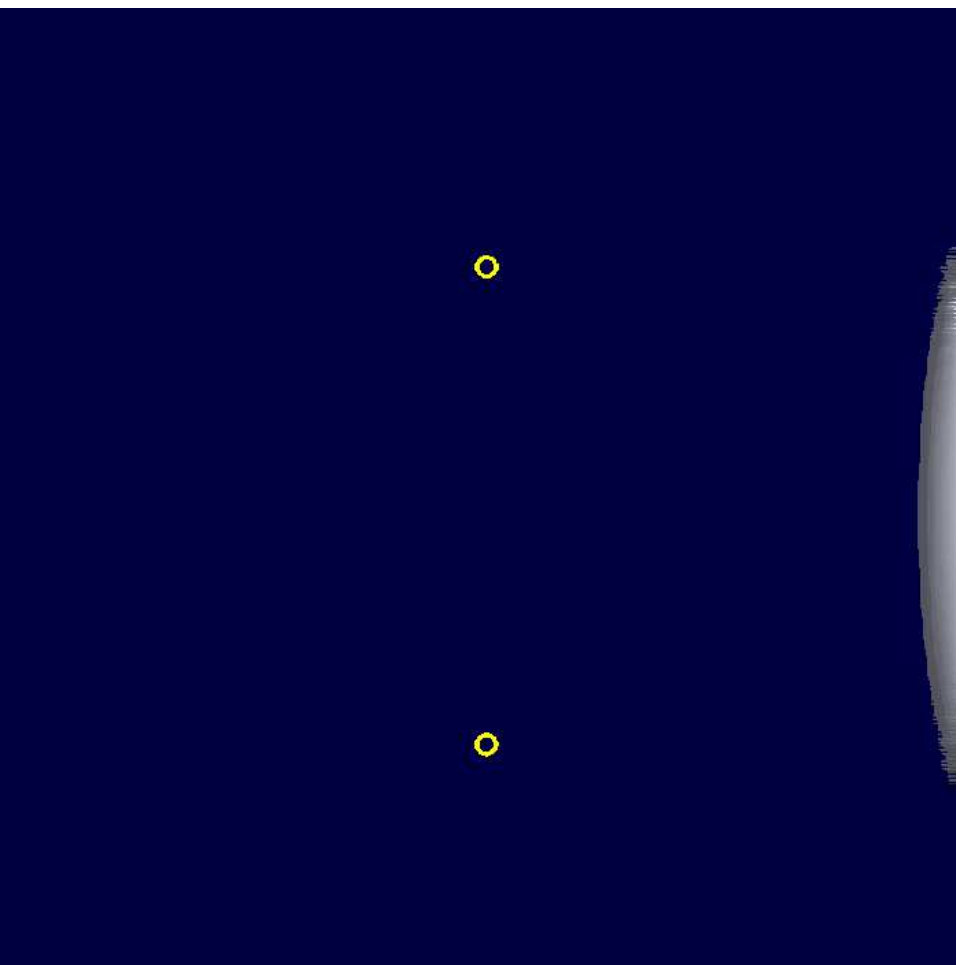}}}
\caption{\label{fig:Wave_two_solenoids} Six stages of the motion of
the wave packet between two infinite parallel solenoids.
Small yellow circles represent the positions and the sizes of the solenoids.
The animation can be accessed on-line \cite{Simi09b}.}
\end{figure}

The motion of the wave packet between the solenoids, in the plane
normal to the orientation of the solenoids, is shown in Figure \ref{fig:Wave_two_solenoids}. 
The wave packet was initially positioned away from center of the solenoids. The dynamics was 
studied for two initial momenta, $p_{1} = 0.53 \; MeV/c$  or $p_{1} = 0.64 \; MeV/c$. 
The solenoids were separated by a distance of $2a=0.1 \; nm$ and the magnitude of the magnetic flux 
inside  each of  them was $\Phi=5.2 \times 10^{-14} \; Wb$.

\begin{figure}
\centering
\vspace*{- 6. cm}
{\scalebox{.75}{\includegraphics{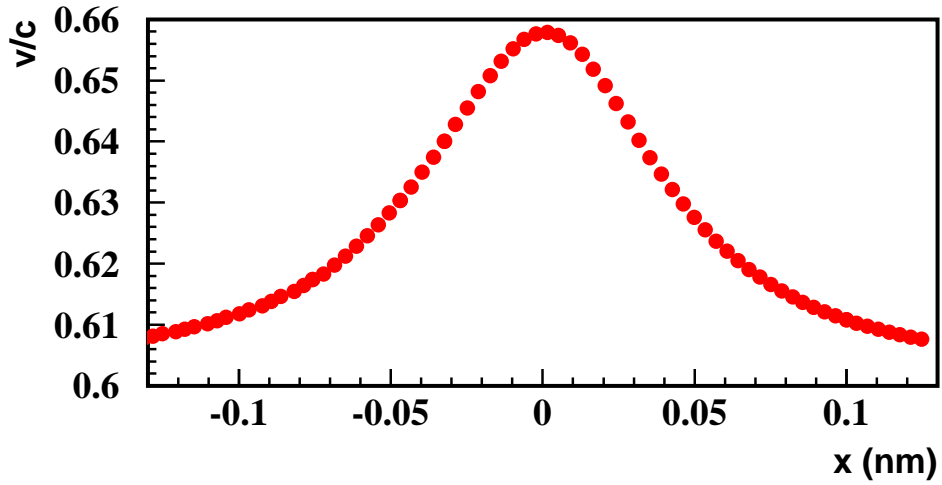}}}
\vspace*{- 7.5 cm}
\caption{\label{fig:two_solenoids_velocity} The velocity of a wave packet 
normalized to the speed of light as a function of the wave packet position.
The wave packet is moving along a straight line between two infinite parallel solenoids positioned at x=0.} 
\end{figure}

As shown in Figure \ref{fig:Wave_two_solenoids}, and in the related animation, the wave packet 
moved between two solenoids along a straight line, dispersing in time in the direction 
normal to the direction of motion. 
Since for the unperturbed solenoid, there was no classical force acting on the electron, 
one should have expected a constant velocity of the wave packet along the straight line.
This was not the case. The velocity
of the wave packet, shown in Figure \ref{fig:two_solenoids_velocity} as a function of the position,
increased as the packet approached the solenoids and decreased as the packet 
left the solenoids. It is somewhat paradoxical that while there was no
classical force expected to act on the electron, the electron acceleration,
obtained strictly as a solution of the Dirac equation, was not zero.

\begin{figure}
\centering
\vspace*{- 2. cm}
\hspace*{- 1. cm}
{\scalebox{.575}{\includegraphics{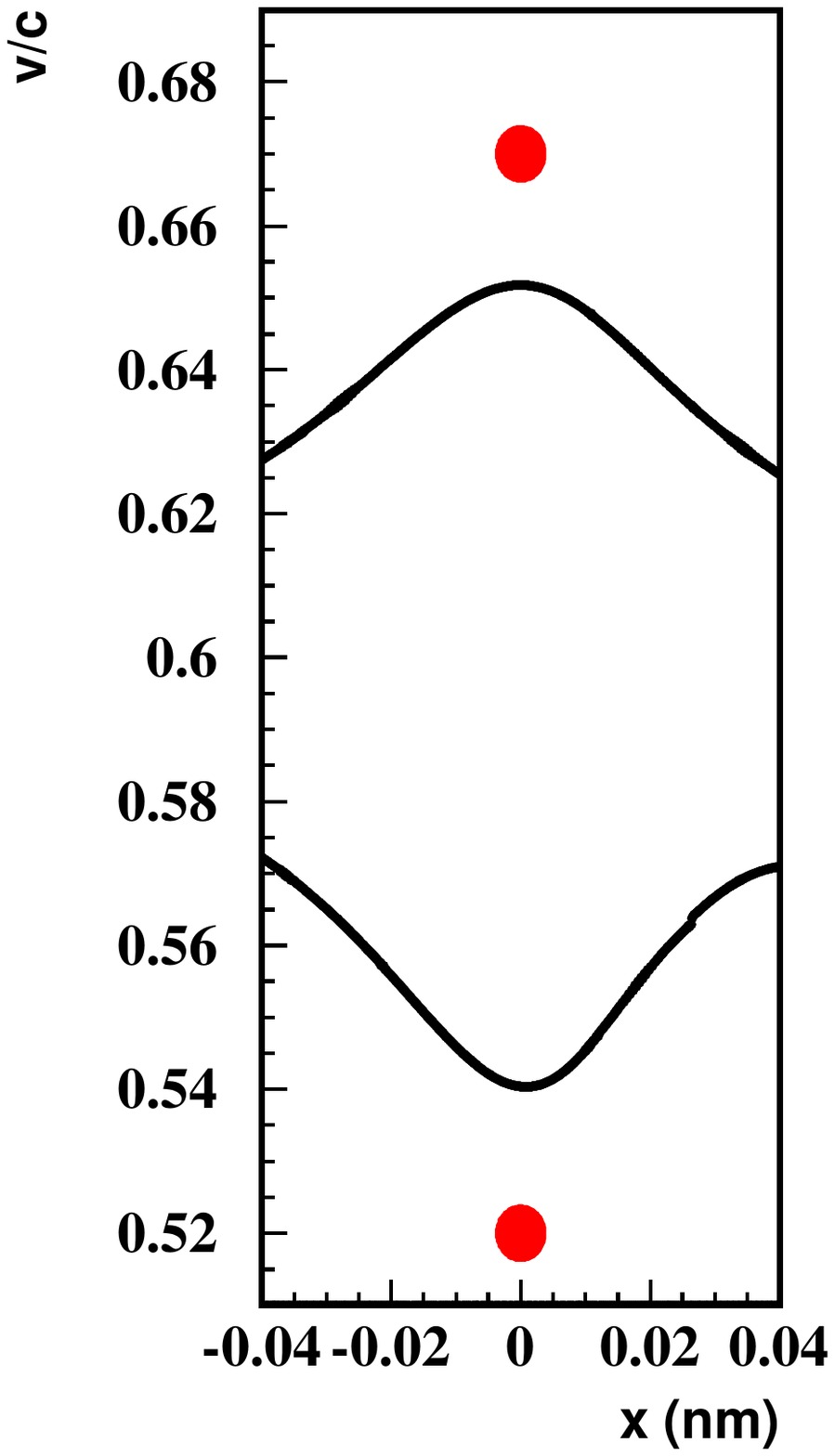}}}\hspace*{- 6.75 cm}
{\scalebox{.575}{\includegraphics{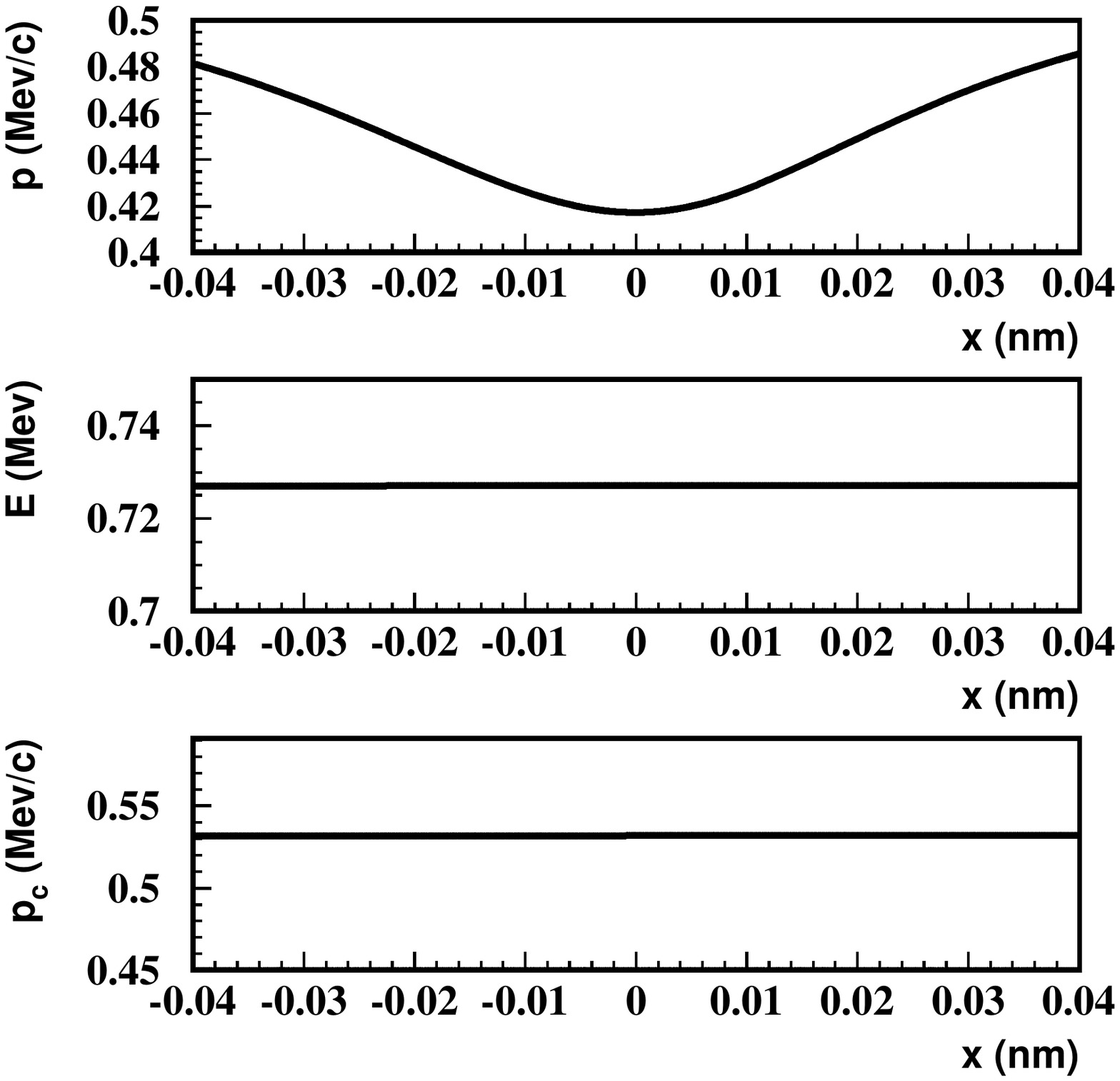}}}
\vspace*{- 4. cm}
\caption{\label{fig:positron_electron_velocities} Left: change of the velocity of a wave packet 
normalized to the speed of light as a function of the wave packet position. Top curve rapresents electron, 
bottom curve positron. The red dots depict the positions of the solenoids at x=0. Right: the top curve shows the 
change in the electron wave packet momentum. Mid and bottom 
lines show that the total wave packet energy and canonical momentum are conserved. The wave packets were 
moving along a straight line between two infinite parallel solenoids separated by 0.072 nm.  The magnitude of the 
magnetic flux inside  each of  them was $\Phi=4.2 \times 10^{-14} \; Wb$. Initial momentum of the wave packets was
0.53 Mev/c.} 
\end{figure}

Extensive studies and verifications of this result were performed. They include the dependence of
the solution of the Dirac equation on particle charge, energy, spin orientation, orientation of the
solenoids, and the strength and the orientation of the magnetic field inside the solenoid.  In addition, the time and
the position dependencies of the particle momentum, energy and canonical momentum were computed. 
The results could be summarized as follows:

\begin{itemize}

\item the change of the velocity of the wave packet did not depend on the particle spin orientation 
relative to the orientation of the solenoids

\item as shown in Figure \ref{fig:positron_electron_velocities}, change of the sign of a charged particle inverted
the order of the change of the velocity, but the velocities converged to the same value at spatial infinity

\item change of the orientation of the magnetic field inside the solenoids had the same effect as change of the
sign of charged particles

\item  as also shown in Figure \ref{fig:positron_electron_velocities}, the mechanical momentum of the particle changed 
during the process, but the total energy and the canonical momentum did not change

\end {itemize}

\section{The dynamics of the charged particle under the conditions of the Aharonov-Bohm effect}

\begin{figure}
\centering
\vspace*{- 3. cm}
{\scalebox{.75}{\includegraphics{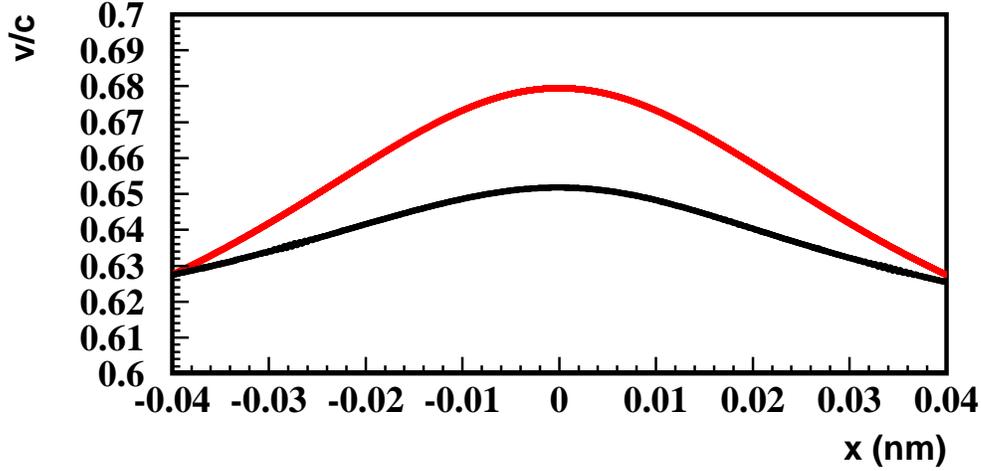}}}
\vspace*{- 11.5 cm}
\caption{\label{fig:theory_two_solenoids_velocity} Higher, red curve, represents the velocity, as a
function of the position, of the relativistic particle moving along a straight line between two 
infinite parallel solenoids, obtained as the solution of  
Eq. (\ref{rel_diff_eq_final}). Lower, black curve, represents the velocity of a wave packet obtained 
from the  solution of the Dirac equation under the
same conditions. The velocities are normalized to the speed of light. The solenoids are positioned at x=0.} 

\end{figure}

The change of velocity was predicted more than 35 years ago by
Liebowitz \cite{Lieb65} and by Boyer \cite{Boy73}, and was proposed as the basis for
the Aharonov-Bohm phase shift. It was shown that if the magnetic
interaction energy of the unperturbed solenoid currents and the passing
charge is assigned to a kinetic energy change of the passing electron,
then the electron has the same qualitative velocity change as found from our
numerical calculation of a wave packet dynamics in the Dirac equation. 
Through
lowest order in the particle-solenoid interaction, the energy conservation
criterion used by Boyer is equivalent to conservation of the passing
charge's canonical momentum $\vec p_{c}=\vec p +q\vec A$. 
Thus 
\begin{eqnarray}
0 &=&\frac{d}{dt}[(p^{2}c^{2}+m^{2}c^{4})^{1/2}+q \vec v \cdot \vec A] \nonumber \\
&=&\frac{\vec p c^{2}}{%
(p^{2}c^{2}+m^{2}c^{4})^{1/2}}\cdot \frac{d\vec p}{dt}+q\vec v \cdot (\vec v \cdot \vec \nabla ) \vec A \nonumber \\
&=&\vec v \cdot \frac{d (\vec p +q\vec A)}{dt}=\vec v \cdot \frac{d\vec p_{c}}{dt}
\label{boyer_no_force}
\end{eqnarray}%
where the term  $q(d\vec v/dt) \cdot \vec A$ was neglected as higher
order in the interaction. This equation implies conservation of the
canonical momentum of the particle. In classical dynamics, the motion of an electron 
can now be  obtained by solving the differential equation
\begin{equation}
{d{(\vec p + q \vec A)} \over dt}=0.
\label{general_no_force}
\end{equation}
Since $A_{x}$ does not explicitly depend on time, for the motion along the x-axis,
Eq. (\ref{general_no_force}) can be written as
\begin{equation}
{d p_{x} \over dt} =  - q {d A_{x} \over dt}=- q {d A_{x} \over dx}{dx \over dt}=- q {d A_{x} \over dx}v_{x}
\label{component_general_no_force}
\end{equation}
In the relativistic case 
\begin{equation}
{d p_{x} \over dt} =\left( 1-{v_{x}^{2} \over c^{2}}\right)^{-{3 \over 2}} m_{0}  {d v_{x} \over dt}. 
\label{rel_momentum}
\end{equation}
Here $m_{0}$ is the rest mass of the electron.  Using Eqs. (\ref{general_no_force}) and (\ref{rel_momentum})
we get the differential equation for the velocity of the electron
\begin{equation}
{d v_{x} \over dt} =- {q \over m_{0}} v_{x} \left( 1-{v_{x}^{2} \over c^{2}}\right)^{3 \over 2}
{d A_{x} \over dx}. 
\label{rel_diff_eq}
\end{equation}
Substituting $A_{x}$ from the Eq.  (\ref{Vect_pot_two_inf_solenoid}) for $z=0$, electron motion along 
the mid-path between two solenoids, the differential equation for the 
velocity of the electron becomes
\begin{equation}
{d v_{x} \over dt} ={{2 q \Phi} \over {\pi m_{0}}} { ax  \over {(x^{2}+a^{2})^{2}}}
v_{x} \left( 1-{v_{x}^{2} \over c^{2}}\right)^{3 \over 2}, 
\label{rel_diff_eq_final}
\end{equation}
where $a$ is half the distance between two parallel infinite solenoids. This differential equation can be
solved numerically. Figure \ref{fig:theory_two_solenoids_velocity} shows the qualitative agreement of the
classical dynamics of the particle obtained from the solution of 
Eq. (\ref{rel_diff_eq_final}) and the quantum dynamics of the corresponding wave packet obtained
from the solution of the Dirac equation.

To conclude this section, we can look at the ongoing question if the  Aharonov-Bohm effect is the result of 
a force changing the  velocity of a particle passing on opposite 
sides of a single infinitely long solenoid, or there is only a quantum-mechanical phase shift 
\cite{Zhu90, Shel98, Keat01,Gronn06,Hors05,Hors07}. Quantum-mechanically
such a question does not exist. The dynamics of the relativistic electron should be obtained only as 
a solution of the time-dependent Dirac equation. The solution
of the Dirac equation shows that the velocity of a wave packet changes even
in the region where the unperturbed solenoid gives zero magnetic field.
Since the change of the velocity depends on the gradient of  the vector potential, the velocity of the 
wave packet passing on the opposite sides of the solenoid will be different.
However, the particle exits the near-solenoid region with the same energy as when it entered,
regardless of the side on which it passed the solenoid.

From the studies presented in this paper, the classical picture of the Aharonov-Bohm effect
consists of the phase shift in the wave function of the particles passing on opposite sides of a solenoid
being attributed to time lag resulting from different evolution of the velocities of the particles. 
A phase shift then results in an interference pattern change.

\section{Conclusion}

In conclusion, the full three-dimensional Finite Difference Time Domain (FDTD)
method was developed to solve the Dirac equation. In this paper, the method was
applied to the dynamics of the electron wave packet in a vector potential
in an arrangement associated with the Aharonov-Bohm effect.
The solution of the Dirac equation showed that the velocity of the electron wave packet 
changed even in the region where the electric and the magnetic fields were zero. 

The solution of the Dirac equation qualitatively agreed
with the prediction of classical dynamics under the assumption that the dynamics were defined by 
the conservation of generalized or canonical momentum. 

The studies in this paper help establishing a picture of the Aharonov-Bohm effect as 
the interference pattern resulting from the phase shift in the wave function of the particles passing 
on opposite sides of a solenoid attributed to a time lag resulting from different evolution 
of the velocities of the particles.

\section*{Acknowledgments}

I would like to thank Prof. Timothy Boyer for all the help, long discussions and encouragements. 
I would also like to thank Profs. Dentcho Genov, B. Ramu Ramachandran, Lee Sawyer, Ray Sterling and Steve Wells 
for useful comments.
Also, the use of the high-performance computing resources provided by Louisiana Optical
Network Initiative (LONI; www.loni.org) is gratefully acknowledged.

\end{document}